\documentclass[reprint,prl,aps, floatfix]{revtex4-1}
\usepackage[utf8]{inputenc}
\usepackage[T1]{fontenc}
\usepackage{graphicx}
\usepackage{xcolor}
\usepackage{dcolumn}
\usepackage{bm}
\usepackage{amssymb}
\usepackage{braket}
\usepackage{color,amsmath}
\usepackage{comment}
\usepackage{gensymb}
\usepackage{soul,xcolor} %use \st{text} to achieve strikethrough
\setstcolor{red} %set colour of strikethrough to red
\usepackage{epstopdf}
\usepackage{hhline}
\usepackage{tabularx}
\usepackage{xspace}
\usepackage{layouts}
\usepackage{lineno} %line numbers for submissionhttps://www.overleaf.com/project/655bc9f3d3ef4a33711ad0e1
\usepackage{xr} %allows referencing between documents
\usepackage{amsfonts}
\usepackage{bbold}
\usepackage{placeins}

\newcolumntype{C}[1]{>{\centering\arraybackslash}m{#1}}
\newcolumntype{N}{@{}m{0pt}@{}}

\begin{document}
% \linenumbers

\title{Superconductivity from dual-surface carriers in rhombohedral graphene}

\author{Manish Kumar$^{1*}$}
\author{Derek Waleffe$^{1*}$} 
\author{Anna Okounkova$^{1*}$}
\author{Raveel Tejani$^{2,3}$}
\author{V\~o Ti\'{\^e}n Phong$^{4,5}$}
\author{Kenji Watanabe$^{6}$}
\author{Takashi Taniguchi$^{7}$}
\author{Cyprian Lewandowski$^{4,5}$}
\author{Joshua Folk$^{2,3\dagger}$}
\author{Matthew Yankowitz$^{1,8\dagger}$}

\affiliation{$^{1}$Department of Physics, University of Washington, Seattle, Washington, 98195, USA}
\affiliation{$^{2}$Department of Physics and Astronomy, University of British Columbia, Vancouver, British Columbia, V6T 1Z1, Canada}
\affiliation{$^{3}$Quantum Matter Institute, University of British Columbia, Vancouver, British Columbia, V6T 1Z1, Canada}
\affiliation{$^{4}$National High Magnetic Field Laboratory, Tallahassee, Florida, 32310, USA}
\affiliation{$^{5}$Department of Physics, Florida State University, Tallahassee, Florida, 32306, USA}
\affiliation{$^{6}$Research Center for Electronic and Optical Materials, National Institute for Materials Science, 1-1 Namiki, Tsukuba 305-0044, Japan}
\affiliation{$^{7}$Research Center for Materials Nanoarchitectonics, National Institute for Materials Science, 1-1 Namiki, Tsukuba 305-0044, Japan}
\affiliation{$^{8}$Department of Materials Science and Engineering, University of Washington, Seattle, Washington, 98195, USA}
\affiliation{$^{*}$These authors contributed equally to this work.}
\affiliation{$^{\dagger}$jfolk@physics.ubc.ca (J.F.) and myank@uw.edu (M.Y.)}

\maketitle

\textbf{Intrinsic rhombohedral graphene hosts an unusual low-energy electronic wavefunction, predominantly localized at its outer crystal faces with negligible presence in the bulk. Increasing the number of graphene layers amplifies the density of states near charge neutrality, greatly enhancing the susceptibility to symmetry-breaking phases. Here, we report superconductivity in rhombohedral graphene arising from an unusual charge-delocalized semimetallic normal state, characterized by coexisting valence- and conduction-band Fermi pockets split to opposite crystal surfaces. In octalayer graphene, the superconductivity appears in five apparently distinct pockets for each sign of an external electric displacement field ($D$). In a moir\'e superlattice sample where heptalayer graphene is aligned on one side to hexagonal boron nitride, two pockets of superconductivity emerge from a single sharp resistive feature. At higher $D$ the same resistive feature additionally induces an $h/e^{2}$-quantized anomalous Hall state at dopings near one electron per moir\'e unit cell. Our findings reveal a novel superconducting regime in multilayer graphene and create opportunities for coupling to nearby topological states.}

Chiral stacking of graphene in the metastable rhombohedral (ABC\ldots) configuration gives rise to unique low-energy electronic properties characterized by flat bands and non-trivial topology~\cite{Xiao2011,Guinea2006,Min2008,Koshino2009,Slizovskiy2019}. Much of the existing work on this material has focused on novel ground states that emerge when a sizable interlayer displacement field ($D$) opens a gap at the charge neutrality point (CNP)~\cite{Shi2020}. Within this gapped regime, the enhanced density of states near the valence and conduction band edges promotes superconductivity~\cite{Zhou2021,Zhou2022,Zhang2023,Holleis2025,Patterson2025,Choi2025,Zhang2025,Yang2025,Morisette2025SC,Li2024,Han2024}, along with spontaneous spin and/or valley polarization in line with the Stoner criterion~\cite{Zhou2021_2,Zhou2022,Han2024Nanotech,Liu2024,Zhou2024}. Recently, superconductivity with compelling signatures of chiral ordering has been observed at small electron doping of the flat conduction band edge in four- to six-layer samples~\cite{Han2024,Morisette2025SC}. When a superlattice potential is further introduced by close rotational alignment of graphene with hexagonal boron nitride (hBN), many-body gaps open at commensurate filling factors of the moir\'e unit cell. Over a similar range of doping and displacement field that hosts the putative chiral superconductivity in misaligned samples, moir\'e devices instead exhibit integer and fractional quantum anomalous Hall (IQAH and FQAH) states~\cite{Lu2024,Lu2025,Waters2025,Xie2025,Ding2025}.

%%%%%%%%%%%%%%%%%%%%%%%%%%%%%%%%%%%%%%%%%%%%%%%%%%%%%%%%%%%%%%%%%%%%%
\begin{figure*}[t]
\includegraphics[width=\textwidth]{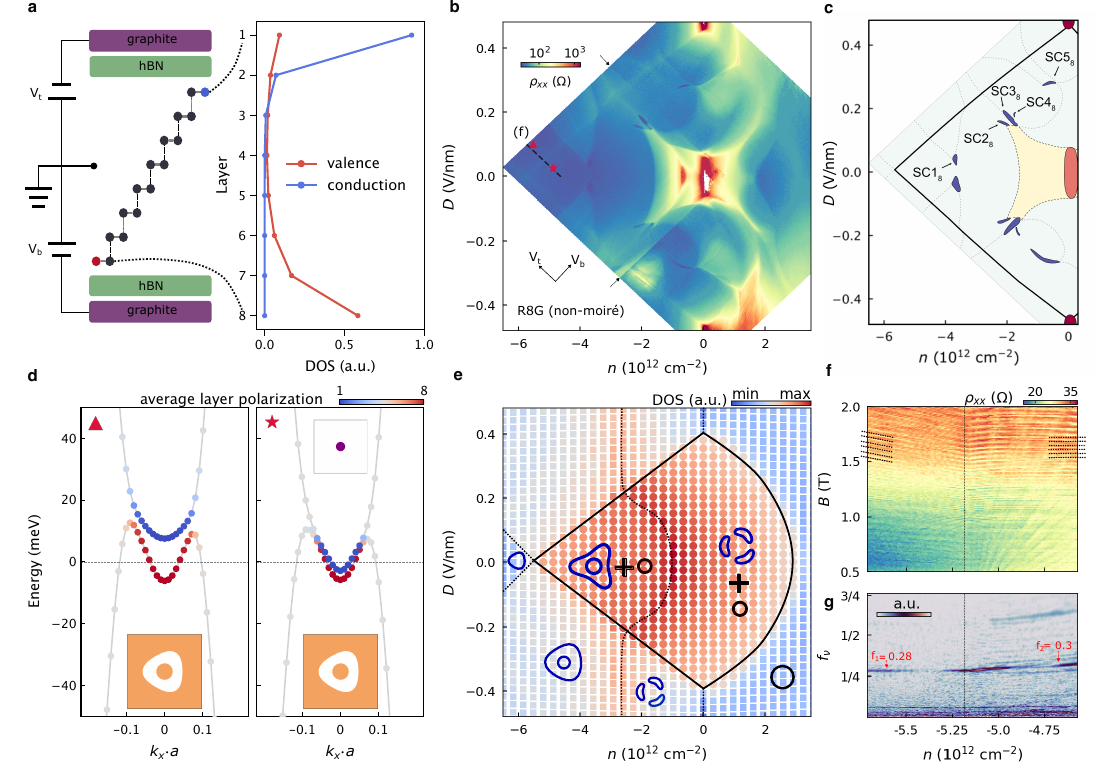} 
\caption{\textbf{Semimetallic properties of rhombohedral graphene and transport in an R8G device.}
\textbf{a}, (left) Schematic of a unit cell of octalayer rhombohedral graphene. Red and blue colors denote the atoms that do not form dimers. The R8G is encapsulated by hBN on both sides, with graphite top and bottom gates. (Right) Calculated density of states resolved by layer number for the valence and conduction band at $n=-1.95\times \mathrm{10^{12}\ cm^{-2}}$ and $D=0.178$~V/nm.
\textbf{b}, Measurement of $\rho_{xx}$ versus $n$ and $D$ in a R8G sample without hBN alignment. Arrows denote measurement artifacts (along fixed $V_t$ and $V_b$) arising from optimizing the silicon gate bias.
\textbf{c}, Schematic denoting some of the key transport features seen in \textbf{(b)}. Band insulators (correlated insulator) at the CNP are denoted in red (dark orange). The light orange indicates a symmetry-broken metallic phase adjacent to the CNP. Dashed and solid lines trace the positions resistive bump features. Blue pockets correspond to superconductivity.
\textbf{d}, Calculated band structure at $n=-5.52\times \mathrm{10^{12}\ cm^{-2}}$ and $D=0.1$~V/nm (left) and $n=-4.87\times \mathrm{10^{12}\ cm^{-2}}$ and $D=0.026$~V/nm (right). The color coding indicates the average layer polarization of each state. The top (bottom) inset shows the constant energy contour of the conduction (valence) band at the Fermi energy ($E=0$). White coloring indicates empty states. $a=2.46$~{\AA} is the graphene lattice constant.
\textbf{e}, Map of the calculated DOS at $E_F$. Solid and dashed lines trace out local maxima or abrupt steps in the DOS, and separate regions of parameter space with distinct Fermi surface topologies (shown in blue and black for the valence and conduction bands, respectively). The solid black curve traces the crossover between the band-overlapping (circular data points) and band-isolated (square data points) regions, and can be compared directly to the black lines in \textbf{(c)}.
\textbf{f}, Landau fan diagram measured by sweeping $V_t$ and fixed $V_b=-6.0$~V, corresponding to the black dashed line in \textbf{(b)}.
\textbf{g}, Fourier transform of the data in \textbf{(f)}.
}
\label{fig:1}
\end{figure*}
%%%%%%%%%%%%%%%%%%%%%%%%%%%%%%%%%%%%%%%%%%%%%%%%%%%%%%%%%%%%%%%%%%%%%

The regime of small interlayer potentials, where the valence and conduction bands instead overlap at the Fermi level to form a semimetal, also hosts a large density of states yet remains far less explored. Substantial charge doping and displacement field are required to move the Fermi energy out of this semimetallic regime, creating a sizable area of parameter space in which rhombohedral multilayer graphene acts as a strongly correlated semimetal. Figure~\ref{fig:1}a shows a rhombohedral octalayer graphene (R8G) device schematic next to a calculation of the layer-resolved density of states (DOS) at a doping of $n=-1.95\times10^{12}\,\mathrm{cm}^{-2}$ and $D=0.178$~V/nm. The conduction band is exponentially localized on the top layer. The valence band is predominantly localized on the bottom layer, but decays more slowly into the bulk before showing a small secondary peak on the bottom layer. This behavior can be understood in analogy with the topological edge states of the Su–Schrieffer–Heeger (SSH) chain model~\cite{Su1979,Xiao2011,Guinea2006,Shi2020}, and demonstrates the propensity for charges in rhombohedral graphene to occupy both surfaces simultaneously with minimal density in the central layers. The bifurcated charge distribution is distinct from that of conventional semimetals, in which electron- and hole-like carriers exhibit substantial spatial overlap. 

Our main observation is that this distinctive semimetallic normal state hosts pockets of superconductivity, for which the distribution of the wavefunction along the $c$-axis emerges as a crucial new degree of freedom mediating pairing. In a moir\'e superlattice sample, the superconductivity emerges from a feature associated with the conduction band in the semimetallic region at small $D$, which then evolves to an integer anomalous Hall state at large $D$ with transverse resistance quantized to $\pm h/e^2$.

\medskip\noindent\textbf{Dual-surface charge distribution}

We first establish the phase boundaries separating the semimetallic region of rhombohedral graphene from regions where the Fermi level crosses only through either the valence or conduction band. We do this by comparing transport measurements with band-structure and Hartree–Fock calculations (full details provided in the Methods and Supplementary Information). Although we focus on R8G, our discussion generalizes to a broad range of layer numbers. 

Figure~\ref{fig:1}b shows a measurement of the longitudinal resistivity ($\rho_{xx}$) from an R8G sample, acquired by independently varying the top- and bottom-gate voltages, $V_{t}$ and $V_{b}$, and converted to $n$ and $D$ through a standard linear mapping (see Methods and Extended Data Fig.~\ref{fig:ggmap7L_8L}). The data reproduce several features familiar from earlier studies of rhombohedral graphene, including band insulators at the CNP at large $D$, a correlated insulator at the CNP centered around $D=0$~\cite{Liu2024,Han2024Nanotech,Myhro2018,Zhang2011,Velasco2012}, and a surrounding region of elevated resistance that has been associated with a symmetry-broken phase~\cite{Han2023Nature}. Figure~\ref{fig:1}c shows a schematic diagram of these and other key features.

Separately, we use band structure calculations (e.g., Fig.~\ref{fig:1}d) to visualize layer polarization of the bands and to generate a map of the DOS at the Fermi energy (Fig.~\ref{fig:1}e). Crucially, each point in this map is obtained from a self-consistent electrostatic calculation, enabling an accurate conversion to the experimentally relevant $n$ and $D$ values~\cite{Kolar2025} (see Methods). The calculation predicts a contiguous region surrounding the charge neutrality point and centered about $D=0$ in which the Fermi level crosses through both the valence and conduction bands. The boundaries of this region are denoted by the solid black outline atop the map in Fig.~\ref{fig:1}e. Inside this region, Fermi surfaces for the valence band (blue) coexist with those for the conduction band (black). Outside, representative sketches illustrate changing topology of the single Fermi surfaces with $n$ and $D$.

The calculation reproduces several salient features of the experiment. Notably, the transport map exhibits diagonal ``jets'' of weakly enhanced $\rho_{xx}$ that emerge upon opening the large-$D$ gap at charge neutrality. These jets follow trajectories of nearly fixed $V_t$ or $V_b$ for $n<0$, denoted by the solid black lines in Fig.~\ref{fig:1}c, that closely correspond to the equivalent segments of the outline in the DOS map. We thus interpret these jets as enclosing the region of $n-D$ space in which the R8G is a semimetal. 

%%%%%%%%%%%%%%%%%%%%%%%%%%%%%%%%%%%%%%%%%%%%%%%%%%%%%%%%%%%%%%%%%%%%%
\begin{figure*}[t]
\centering
\includegraphics[width=0.9\textwidth]{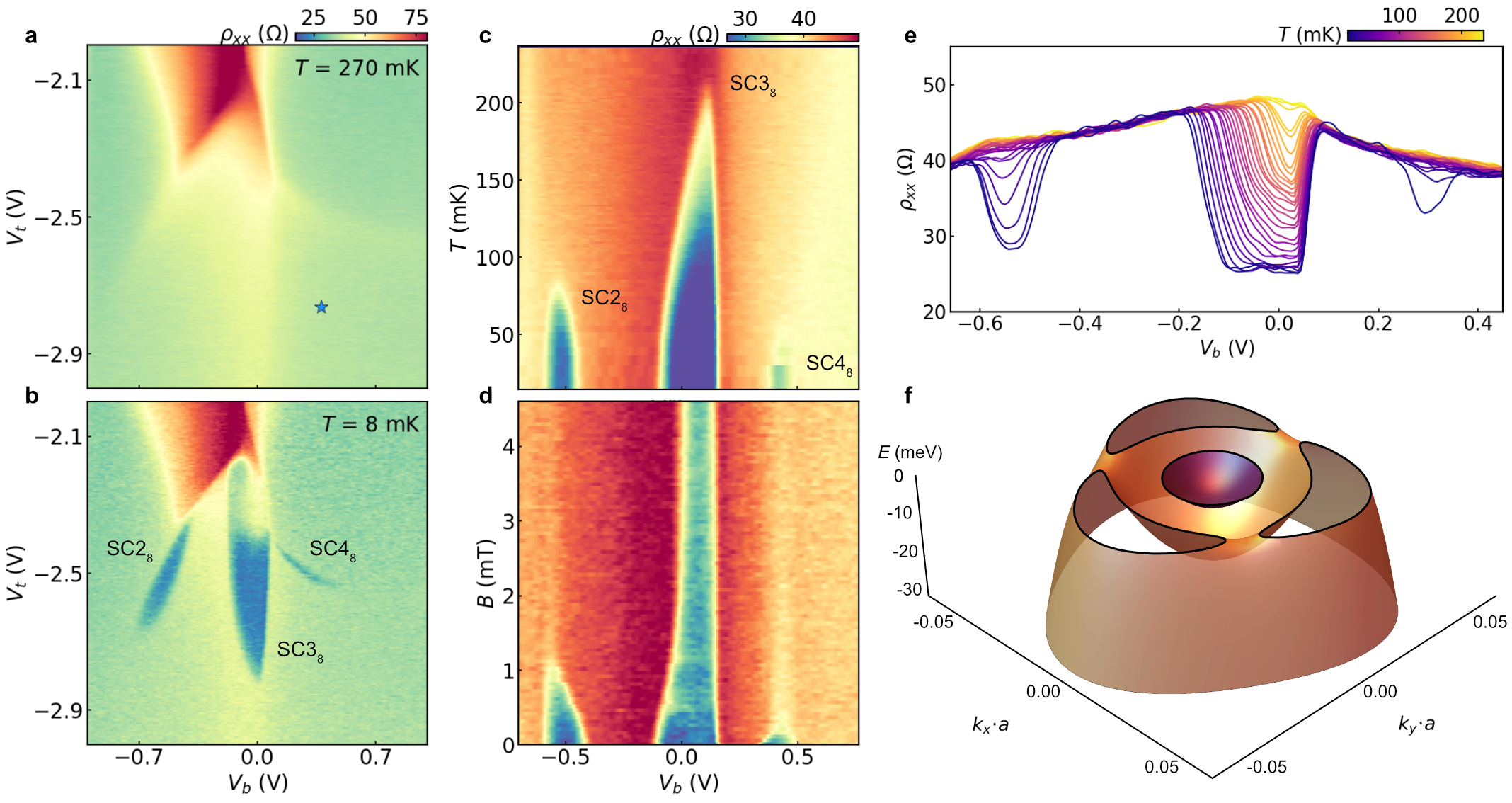} 
\caption{\textbf{Superconductivity in the dual-surface carrier regime of R8G.}
\textbf{a}, Zoomed-in map of $\rho_{xx}$ versus $V_t$ and $V_b$ at $T=270$~mK and $B=0$.
\textbf{b}, The same map at $T=8$~mK. 
\textbf{c}, Measurements of $\rho_{xx}$ versus $V_b$ and $T$ at fixed $V_t=-2.55$~V.
\textbf{d}, Similar measurement versus $B$.
\textbf{e}, Line trances of $\rho_{xx}$ versus $V_b$ from \textbf{(c)} at various selected $T$.
\textbf{f}, Calculated Hartree-Fock band structure, $E$ versus $k_x$ and $k_y$, at $n=-1.96\times10^{12}$~cm$^{-2}$ and $D=0.203$~V/nm (corresponding to the position of the blue star in \textbf{(a)}. The conduction (valence) band is shown in purple (orange). The Fermi surface is denoted at $E_F=0$ with a solid black outline. 
}
\label{fig:2}
\end{figure*}
%%%%%%%%%%%%%%%%%%%%%%%%%%%%%%%%%%%%%%%%%%%%%%%%%%%%%%%%%%%%%%%%%%%%%

To support this interpretation, we examine a Landau fan diagram taken along the black dashed line in Fig.~\ref{fig:1}b, in which magnetic field ($B$) and $V_{t}$ are swept at fixed $V_{b}$ (Fig.~\ref{fig:1}f). The quantum oscillations (QOs) at very negative $n$ follow the Středa formula, projecting to $n=0$ at $B=0$. Upon crossing the jet at $n=-5.18\times 10^{12}$, those oscillations become nearly horizontal---almost insensitive to $V_{t}$, as if screened---and a new series of features appears, dispersing toward the same $V_{t}$ value that corresponds to the position of the jet. A Fourier transform in the unscreened region (left half of the graph) shows a dominant frequency ($f_v$) slightly above $1/4$ (Fig.~\ref{fig:1}g), consistent with a four-fold--degenerate annular Fermi surface (see Methods). This frequency is nearly constant at large negative $n$ but increases upon crossing the jet, consistent with the simultaneous filling of an additional conduction band Fermi pocket. These observations agree well with the band structure calculations corresponding to points of the Landau fan on either side of the jet (Fig.~\ref{fig:1}d), which show four-fold degenerate isolated and overlapping valence and conduction bands at the Fermi energy, respectively. 

\medskip\noindent\textbf{Superconductivity in semimetallic R8G}

We find ten low-resistance pockets within the band-overlapping region that, as we show below, correspond to the emergence of superconductivity (Figs.~\ref{fig:1}b-c). Five pockets appear for each sign of $D$, labeled as SC1$_8$ to SC5$_8$ for $D>0$ (the subscript denotes the total number of graphene layers). We henceforth restrict our focus to the three pockets that appear near $V_b=0$ (SC2$_8$, SC3$_8$, and SC4$_8$), and analyze the other two pockets (SC1$_8$ and SC5$_8$) and $D<0$ counterparts in Extended Data Figs.~\ref{fig:T_B_comparison}-\ref{fig:8L_negD_SC}. Figures~\ref{fig:2}a,b show detailed $\rho_{xx}$ maps of the region surrounding these pockets, taken at $T=270$~mK and $8$~mK, respectively. Contours of slightly elevated resistance, seen at higher temperature, develop into three pockets of deeply suppressed $\rho_{xx}$ at base temperature. Figures~\ref{fig:2}c,d show $\rho_{xx}$ versus $T$ and $B$ for the three pockets (taken by sweeping $V_b$ at fixed $V_t)$, exhibiting three dome-like structures consistent with superconductivity. The critical temperatures are small, with $T_c \approx 20$~mK to $170$~mK (Fig.~\ref{fig:2}c and Extended Data Table~\ref{tab:SC_summary}). The corresponding critical fields are similarly low, with $B_c$ on the order of a few millitesla or less. Measurements of d$V$/d$I$ versus dc bias current $I_{dc}$ and $B$ also show distinct nonlinearities characteristic of superconductivity (see Extended Data Fig.~\ref{fig:8L_allSC)}). However, the resistance does not reach zero in any of the five pockets, and d$V$/d$I$ measurements show unexpected peaks at $I_{dc}=0$ (see further discussion in Methods). Although we do not know the origins of either feature, we note that a non-zero saturation of resistance is commonly reported in low-$T_c$ pockets from thinner rhombohedral graphene samples~\cite{Zhou2021,Patterson2025,Holleis2025,Choi2025,Zhang2025,Yang2025}.

It is notable that all superconducting pockets appear along a nearly fixed value of $V_t$ ($V_b$) for $D>0$ ($D<0$). We speculate that this may arise because charge carriers in the normal state reside mainly on the two outer crystal surfaces and are separated by a low-DOS bulk (Fig.~\ref{fig:1}a). The system thus behaves as two nearly isolated electron gases, held at a common chemical potential by the small fraction of layer-delocalized states. Each gate couples most strongly to the surface closest to it, with minimal cross-influence due to screening~\cite{Kolar2025}, locking features to nearly fixed $V_{t}$ or $V_{b}$ rather than along the usual $n-D$ axes. This screening behavior can be directly understood from the Hartree-Fock band structure calculations (e.g., Fig.~\ref{fig:1}d), in which the color scale denotes the average layer polarization of each state. States near the K and K' points ($k=0$) predominantly reside on the outer surfaces and can screen their neighboring gates when populated, whereas states at larger $|k|$ are layer-delocalized and leak across the bulk of the rhombohedral flake. As argued in Ref.~\citenum{Kolar2025} and the Supplementary Information, this spatial structure of the wavefunction naturally gives rise to a single-gate tracking behavior, exemplified by the diagonal semimetallic phase boundaries for $n<0$ in Fig.~\ref{fig:1}e. In this context, the five pockets of superconductivity apparently correspond to a nearly constant doping of the surface carrying the conduction band, and are modulated by doping the valence band, which resides primarily on the opposite surface.

To the best of our knowledge, there is no other known material system in which superconductivity forms from such a charge-bifurcated distribution, with valence and conduction bands polarized to opposite surfaces. An open question is whether superconducting gaps form in both the valence and conduction bands simultaneously, and if so, whether the gaps are comparable in magnitude, or if one band instead induces superconductivity in the other through a proximity effect. At minimum, the gate-tracking characteristics described above indicate that both bands play a role in the superconductivity. This observation invites new considerations of how the extended spatial distribution of the wavefunction is manifest in attempts to theoretically describe pairing in this system. Our band structure calculations are especially noteworthy for values of $n$ and $D$ around where SC2$_8$, SC3$_8$, and SC4$_8$ appear, in which a local maximum in the DOS coincides with a Lifshitz transition in the valence band and a small simply-connected Fermi pocket from the conduction band (Figs.~\ref{fig:1}e and~\ref{fig:2}f). We emphasize the distinction from other crystalline graphene systems, where superconductivity likewise tracks features potentially associated with a high density of states, but without the additional interplay provided by the charge-delocalized semimetallic normal state. Further work will be needed to understand which features of each band promote superconductivity.

%%%%%%%%%%%%%%%%%%%%%%%%%%%%%%%%%%%%%%%%%%%%%%%%%%%%%%%%%%%%%%%%%%%%%
\begin{figure*}[t]
\includegraphics[width=\textwidth]{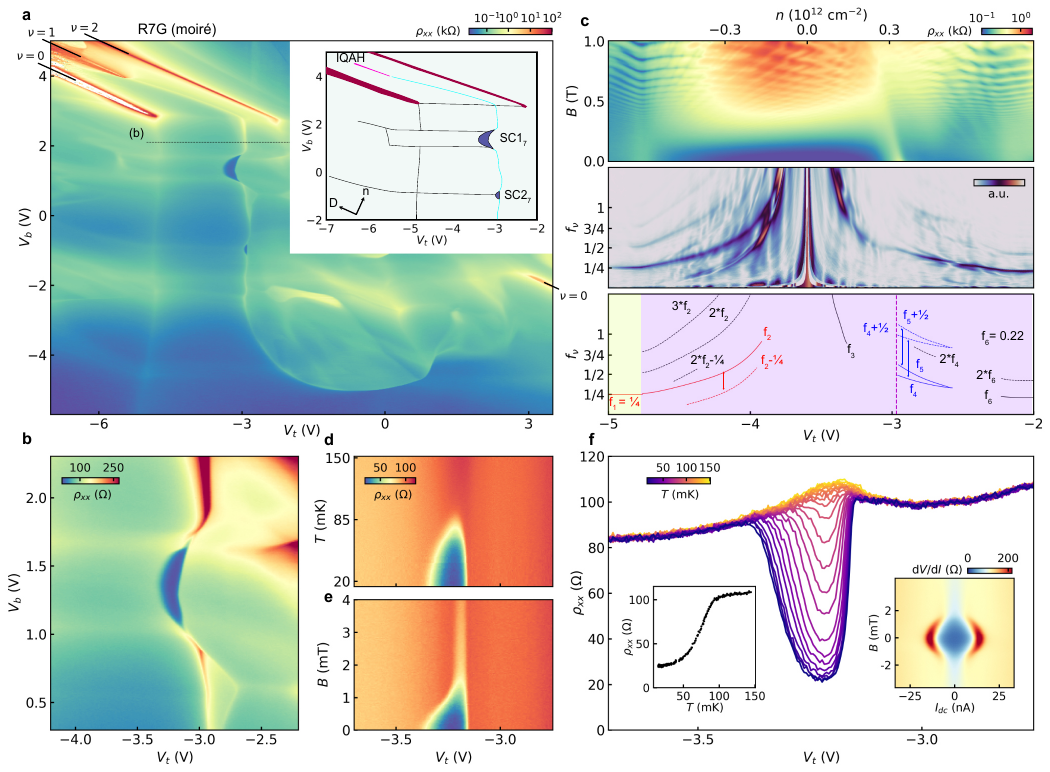} 
\caption{\textbf{Superconductivity in a moir\'e R7G device.}
\textbf{a}, Map of $\rho_{xx}$ versus $V_t$ and $V_b$ taken at base temperature and $B=0$. (inset) Schematic denoting some of the key transport features. Trivial (Chern) insulators at integer $\nu$ are shown in red (pink). Black and blue curves denote resistance bump features. Dark blue pockets correspond to superconductivity.
\textbf{b}, Zoomed-in map of $\rho_{xx}$ around the region of SC1$_7$.
\textbf{c}, (top) Landau fan taken by sweeping $V_t$ at fixed $V_b=2.1$~V, along the black dashed line in \textbf{(a)}. (middle) Fourier transform of the Landau fan. (bottom) Schematic denoting the dominant frequencies seen in the FFT. Background colors denote the band-isolated (yellow) and band-overlap regions (purple).
\textbf{d}, Measurements of $\rho_{xx}$ versus $V_t$ and $T$ at fixed $V_b=1.4$~V.
\textbf{e}, Similar measurement versus $B$.
\textbf{f}, Line trances of $\rho_{xx}$ at various selected $T$. (left inset) Measurement of $\rho_{xx}$ versus $T$. (right inset) Measurement of d$V$/d$I$ versus $I_{dc}$ and $B$. Measurements in both insets are taken at the optimal doping of SC1$_7$ ($V_t=-3.22$~V and $V_b=1.4$~V).
}
\label{fig:3}
\end{figure*}
%%%%%%%%%%%%%%%%%%%%%%%%%%%%%%%%%%%%%%%%%%%%%%%%%%%%%%%%%%%%%%%%%%%%% 

\medskip\noindent\textbf{Superconductivity in a moir\'e R7G sample}

The complementary role of layer-polarized conduction and valence bands in inducing superconductivity is even more clear in measurements of a sample of rhombohedral heptalayer graphene (R7G) that is aligned with hBN, giving a 13.5~nm moir\'e superlattice. Figure~\ref{fig:3}a shows a map of $\rho_{xx}$ as a function of $V_b$ and $V_t$ for the R7G sample at $B=0$, with key features summarized schematically in the inset. Compared with the misaligned R8G device, this sample lacks the correlated insulator and symmetry-broken metal near $n=D=0$, but at larger $D$ develops correlated insulating states at commensurate fillings ($\nu$) of the moir\'e conduction band (see also Extended Data Fig.~\ref{fig:ggmap7L_8L}). We observe two distinct superconducting pockets (SC1$_7$ and SC2$_7$) emerging out of a sharp resistance bump feature that tracks fixed $V_t$ (shown in light blue in the schematic in Fig.~\ref{fig:3}a). The more robust pocket, SC1$_7$, is bounded by two perpendicular bumps that closely track fixed $V_b$ (Fig.~\ref{fig:3}b). This bounding is similar to features associated with the SC5$_8$ pocket in the R8G sample (Extended Data Fig.~\ref{fig:R7G_R8G_SC_compare}). The weaker pocket, SC2$_7$, is also directly connected to a perpendicular resistance bump. Both SC1$_7$ and SC2$_7$ lie within the band-overlapping regime, as determined by analysis of Landau fans taken across the jet emerging from the large-$D$ insulating state at the CNP (Fig.~\ref{fig:3}c; additional data and analysis in Methods, Extended Data Fig.~\ref{fig:Fermiology7LVt-4V} and Supplementary Information Fig.~\ref{fig:LFs_SC1_SC2}). In particular, in Fig.~\ref{fig:3}c we see a dispersing sequence of QOs with $f_v=1/4$ for $V_t<-4.75$~V and screening of those QOs with a corresponding rapid increase of $f_v$ for $V_t>-4.75$~V, suggesting that the jet from the CNP defines the band-overlapping to -isolated region, analogous to the behavior seen in the R8G sample.

%%%%%%%%%%%%%%%%%%%%%%%%%%%%%%%%%%%%%%%%%%%%%%%%%%%%%%%%%%%%%%%%%%%%%
\begin{figure*}[t]
\includegraphics[width=\textwidth]{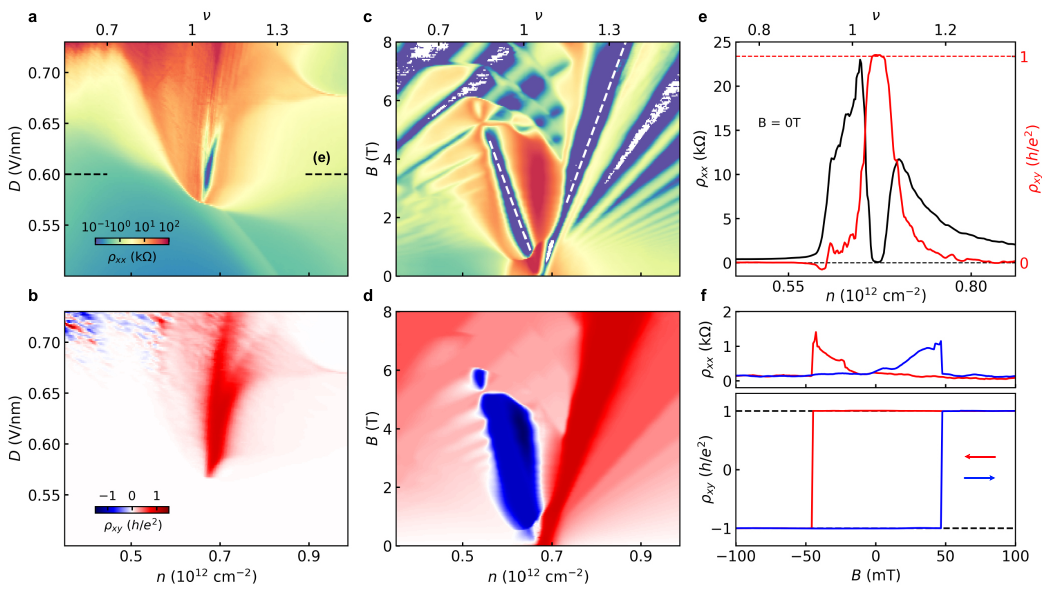} 
\caption{\textbf{Integer quantum anomalous Hall state near $\nu=1$ in the moir\'e R7G device.}
\textbf{a}, Map of $\rho_{xx}$ versus $\nu$ and $D$ at $B=0$.
\textbf{b}, Similar map of $\rho_{xy}$, antisymmetrized at $|B|=100$~mT.
\textbf{c}, Landau fan of $\rho_{xx}$ taken at $D=0.60$~V/nm (marked by the dashed lines in \textbf{(a)}). White dashed lines indicate the Streda trajectories of $C=+1$ and $-1$ states. White colors indicate negative measured resistances.
\textbf{d}, Same for $\rho_{xy}$.
\textbf{e}, Line traces of $\rho_{xx}$ and $\rho_{xy}$ versus $\nu$ at $B=0$ and $D=0.60$~V/nm.
\textbf{f}, Measurement of $\rho_{xx}$ and $\rho_{xy}$ as $B$ is cycled at $\nu=1.06$ and $D=0.60$~V/nm. \textbf{a} and \textbf{e} are taken at $T=9$~mK. All other measurements are taken at $T=100$~mK.
}
\label{fig:4}
\end{figure*}
%%%%%%%%%%%%%%%%%%%%%%%%%%%%%%%%%%%%%%%%%%%%%%%%%%%%%%%%%%%%%%%%%%%%%

Figures~\ref{fig:3}d,e show measurements of $\rho_{xx}$ for SC1$_7$ versus $T$ and $B$ as $V_t$ is swept at fixed $V_b\approx1.40$~V. Both exhibit characteristic domes similar to those seen in R8G. Figure~\ref{fig:3}f displays line traces of $\rho_{xx}$ versus $V_g$ for several selected $T$. There is a small bump in the resistance at higher $T$, which develops into a deep suppression of $\rho_{xx}$ at base temperature. A measurement of $\rho_{xx}$ versus $T$ shown in the left inset highlights the abrupt downturn of the resistance below $\approx 90$~mK, before saturating near 20~Ohms. The d$V$/d$I$ measurements in the right inset of Fig.~\ref{fig:3}f provide additional evidence for superconductivity, showing sharp peaks at a critical current of $\approx 15$~nA and a recovery of nearly linear transport above $B_c$. Analogous measurements for SC2$_7$ are shown in Extended Data Fig.~\ref{fig:SC7_2}.

The $\rho_{xx}$ bump that develops into the two superconducting pockets in R7G seems clearly to be a feature of the conduction band, since it persists into the electron-doped region ($n>0$) at large $D$ when a gap opens at the CNP, and because its trajectory becomes nearly independent of $V_b$ within the dual-surface regime due to screening from coexisting valence band states on the bottom surface. Detailed analysis of the QOs in Landau fan measurements across the bump feature suggests that it separates an isospin-unpolarized metal to the left from a spin-polarized half-metal to the right (Fig.~\ref{fig:3}c), although alternative interpretations remain possible (see Methods). The resistive bump shifts to more negative $V_t$ with increasing $B$ (Fig.~\ref{fig:3}c), further consistent with a spin-polarized state to the right gaining Zeeman energy over the unpolarized state to the left. 

In this picture, SC1$_7$ lies on the boundary between isospin unpolarized and polarized states within the conduction band, and is turned on and off by valence band features determined by charges on the opposite crystal surface. Indeed, in-plane field measurements of the superconductivity indicate an unconventional isospin ordering and will be the subject of future work. SC2$_7$ shares similar characteristics. Both pockets appear only when conduction band states are polarized away from the moir\'e interface, suggesting that superconductivity is disfavored when these states are moir\'e-adjacent. These conclusions are supported by our self-consistent Hartree-Fock simulations, which show that the presence of a moir\'e superlattice disfavors the charge-separated semimetal for displacement fields polarizing the conduction band towards the moir\'e interface (see Supplementary Information for further discussion).

\medskip\noindent\textbf{Chern insulator at large displacement field}

Lastly, we show that the same conduction-band feature that hosts superconductivity in the semimetallic region of the moir\'e R7G device also hosts an integer quantum anomalous Hall state once the bands are separated, with quantization of the Hall resistance to $\rho_{xy}=\pm h/e^2$. Upon increasing $D$ and exiting the semimetallic regime, the resistance bump abruptly becomes responsive to both $V_t$ and $V_b$, consistent with a disappearance of valence-band--mediated screening from the remote gate. Figures~\ref{fig:4}a,b are 2D scans along $n$ and $D$ axes corresponding to the top left corner of Fig.~\ref{fig:3}a, showing the evolution of the resistive bump into an IQAH state when it drifts to $\nu \approx 1$ at $D\gtrsim 0.58$~V/nm. The IQAH state is marked by the simultaneous suppression of $\rho_{xx}$ towards zero (Fig.~\ref{fig:4}a) and quantization of $|\rho_{xy}|$ to $h/e^2$ (Fig.~\ref{fig:4}b). In a Landau fan (Figs.~\ref{fig:4}c,d), the relevant $\rho_{xx}$ and $\rho_{xy}$ features slope with a trajectory consistent with $C=+1$ under the Středa formula, and persist to at least $8$~T. Line traces at $B=0$ (Fig.~\ref{fig:4}e) confirm the vanishing of $\rho_{xx}$ and quantization of $\rho_{xy}$ within a narrow region of $\nu$. Additional sweeps taken at $\nu=1.06$ while cycling $B$ (Fig.~\ref{fig:4}f) show a single hysteresis loop with $\rho_{xy}$ switching between $\pm h/e^2$, providing definitive evidence for the $|C|=1$ IQAH state. The appearance of a IQAH state with a Chern number of $C=1$ at high $D$ is in line with previous measurements on four- to ten-layer rhombohedral graphene~\cite{Lu2024,Lu2025,Waters2025,Xie2025,Xie2025_2}.

Two features in Fig.~\ref{fig:4} extend beyond conventional IQAH phenomenology. First, the Chern insulator is apparently not pinned exactly to $\nu=1$ (see Extended Data Fig.~\ref{fig:7L_IQAH}). In fact, all of the correlated insulators on the moir\'e-remote side shift progressively away from integer $\nu$ as $D$ increases (see Supplementary Information Fig.~\ref{fig:Dcomparison}), underscoring the need to clarify the experimental conditions for opening many-body gaps in thicker rhombohedral samples. Second, the Landau fan reveals an additional $C=-1$ state that emerges abruptly for $B\gtrsim0.6$~T, similar to behavior recently reported in moir\'e R6G~\cite{Xie2025_2} and twisted bilayer-trilayer graphene~\cite{Su2025}. Our Hartree--Fock calculations that allow for isospin symmetry-breaking cannot yet faithfully capture the $|C|=1$ nature of the observed IQAH state, suggesting instead that it should be trivial. We conjecture that this mismatch may be a consequence of our particular choice of non-interacting Hamiltonian parameters, or could point towards a more complex pattern of symmetry breaking (see further discussion in Methods).

\medskip\noindent\textbf{Discussion}

Superconductivity appears widely across crystalline graphene multilayers, but it is unclear whether the underlying pairing mechanism is universal. The unusual charge-delocalized semimetallic normal state we identify introduces a new theoretical scenario in which two spatially separated electronic reservoirs can interact, reshaping the effective pairing interaction. Gate-tracking behavior in both R8G and moir\'e R7G indicates that superconductivity is primarily tied to features in the conduction band, suggesting that the valence band plays a secondary---yet still important---role. Regardless of pairing mechanism, superconductivity requires an attractive interaction to overcome the renormalized Coulomb repulsion. Within the Tolmachev–Anderson–Morel framework for phonon-mediated superconductivity~\cite{tolmachev1961,morelanderson1962}, additional carriers in the valence band could enhance screening and thereby strengthen pairing. Alternatively, the charge-separated geometry resembles ``synthetic'' electronic pairing proposals~\cite{little1964,Ginzburg1964,bardeen1973,Malozovsky1996,Hamo2016,PhysRevB.98.094517,kivelson2017} in which one charge layer mediates an effective attraction in another (e.g., dynamical, excitonic, or Kohn–Luttinger scenarios). Furthermore, the propensity of the crystalline graphene systems for symmetry breaking necessitates consideration of pairing arising from fluctuations of isospin degrees of freedom~\cite{scalapino2012}, in which screening from the additional charge layer would naively suppress Coulomb-driven pairing; however, careful consideration is required.

Estimates of the superconducting coherence length, mean free path, and Fermi wavevectors place the system in the clean, weak-coupling limit, though extracting these parameters is complicated by the coexistence of valence- and conduction-band carriers (see Methods and Extended Data Table~\ref{tab:SC_summary}). If confirmed, this regime renders the pairing problem in rhombohedral graphene particularly amenable to controlled theoretical treatments. Future experiments tracking the evolution of superconductivity in the semimetallic regime as a function of layer number will likely be helpful for unraveling its nature. A recent study of R6G revealed superconductivity in a similar region of parameter space as the SC3$_8$ pocket~\cite{Morisette2025SC}; our measurements extend the presence of superconductivity beyond six-layer samples for the first time, motivating the possibility that it persists into the regime of bulk rhombohedral graphite. Thicker flakes are of particular interest since the larger spatial separation of the two surfaces could sensitively modulate the presence and strength of superconductivity.  

Taken together, our findings raise new questions about superconductivity in graphene multilayers, in which the highly nonuniform charge distribution along the stacking axis appears to play a key role in pairing. They also reveal intriguing connections between superconductivity and topological many-body phases in moir\'e superlattices. The evolution of a sharp resistive bump feature from a superconducting state at low $D$ into an IQAH state at higher $D$ in moir\'e R7G highlights the versatility of this platform for studying the interplay between superconductivity and topology, potentially opening pathways toward engineered topological superconductivity and exotic non-Abelian phases.

\section*{Methods}

\textbf{Device fabrication.} We first identified regions of likely rhombohedral stacking within flakes of exfoliated graphene using an infrared camera mounted on an optical microscope. We then mapped the stacking domains in detail with amplitude-modulated Kelvin probe force microscopy (AM-KPFM) using a Bruker Icon atomic force microscope (AFM) with an SCM-PIT-V2 tip~\cite{Waters2025}. The rhombohedral region of interest was subsequently isolated using a resist-free local anodic oxidation nanolithography process~\cite{Li2018}. To create the van der Waals (vdW) heterostructure, we first prepared the bottom portion of the stack by picking up an hBN and graphite dielectric/back-gate and placing them onto an SiO$_2$ substrate. We then cleaned the surface of the hBN using contact-mode AFM (using a OTESPA-R3 AFM tip and a line spacing of $\approx 100$ nm). We separately prepared the top half of the stack, consisting of graphite/hBN/RG, and deposited it onto the back half. We used standard dry transfer techniques with a polycarbonate (PC) film on a polydimethylsiloxane (PDMS) stamp for all vdW assembly. For the moir\'e R7G sample, straight edges of the RG flake were aligned to those of the hBN.

After stacking, standard device fabrication procedures were employed to create the dual-gated Hall bar device (i.e., reactive ion etching and evaporation of 7/70 nm of Cr/Au, all using poly(meth)acrylate (PMMA) masks patterned by e-beam lithography). Extended Data Fig.~\ref{fig:ggmap7L_8L} shows optical micrographs of the completed devices.

\textbf{Transport measurements.} Transport measurements were carried out in a Bluefors XLD dilution refrigerator with a one-axis superconducting magnet. The moir\'e R7G device was subsequently measured in a Bluefors LD dilution refrigerator equipped with a 3-axis superconducting vector magnet. Unless otherwise specified, measurements were carried out at the nominal base mixing chamber temperature of the fridge ($T=8$~mK, as measured by a factory-supplied RuO$_x$ sensor). Four-terminal lock-in measurements were performed by sourcing a small alternating current between $I_{ac}=100$~pA and $6.67$~nA at a frequency $<40$ Hz, chosen to accurately capture sensitive transport features while minimizing electronic noise. All $\rho_{xx}$ and d$V$/d$I$ measurements are presented after multiplying the raw measured resistance values by the aspect ratio $W/L$, set by the width ($W$) and length ($L$) of the portion of the Hall bar between the voltage probes. In addition, a global bottom gate voltage between $-20$~V and $+20$~V was applied to the Si substrate to improve the contact resistance. The silicon gate voltage was adjusted between different portions of the map shown in Fig.~\ref{fig:1}b, resulting in the artificial boundaries indicated by the black arrows.

The charge carrier density, $n$, and the out-of-plane electric displacement field, $D$, were defined according to $n= \left(C_{\text{b}} V_{\text{b}}+C_{\text{t}} V_{\text{t}}\right) / e$ and $D=\left(C_{\text{b}} V_{\text{b}} - C_{\text{t}} V_{\text{t}}\right) / 2 \epsilon_0$, where $C_{\text{t}}$ and $C_{\text{b}}$ are the top and bottom gate capacitance per unit area and $\epsilon_0$ is the vacuum permittivity. $C_{\rm{t}}$ and $C_{\rm{b}}$ were estimated by fitting the slopes of the quantum Hall states in Landau fan diagrams.

In some Landau fan diagrams, we symmetrized $\rho_{xx}$ and antisymmetrized $\rho_{xy}$ to reduce the effects of geometric mixing, following the relations $\rho_{xx}=\left(\rho_{xx}(B > 0) + \rho_{xx}(B < 0)\right)/2$ and $\rho_{xy}=\left(\rho_{xy}(B > 0) - \rho_{xy}(B < 0)\right)/2$. In measurements of superconductivity versus $B$, we adjusted the nominal value of $B$ by a few millitesla such that the $\rho_{xx}$ data is symmetric about $B=0$, which was necessary due to the small trapped flux in the superconducting magnet coil.

\textbf{Layer number determination.} To determine the number of graphene layers in the misaligned sample, we examine a Landau fan at $D=0$ (Supplementary Information Fig.~\ref{fig:R8G_layernum}). As explained in detail in Ref.~\cite{Shi2020}, at high $B$ the most robust Landau level gap corresponds to a filling factor of $\nu_{LL}=-2N$, where $N$ is the number of graphene layers. In the misaligned sample, we find that the most robust filling factor is $\nu_{LL}=-16$, providing an unambiguous identification of the flake as having eight layers.

In the moir\'e sample, the added complexity in the band structure precludes the application of this technique. In thinner samples, there is additionally a magnetic field driven Chern insulator state arising from the CNP that can be used to determine the layer number~\cite{Waters2025}, however, here we find that in thicker moir\'e samples such a state apparently does not arise. Thus, we rely on optical contrast to determine the layer number. Supplementary Information Fig.~\ref{fig:R7G_layernum} shows an optical micrograph of the flake. By carefully counting monolayer and bilayer steps in the graphene flake, as indicated by their changes in optical contrast, we clearly identify this flake as having seven layers.

\textbf{Determination of moir\'e period and twist angle.} To extract the moir\'e wavelength in the aligned R7G device, we fit the sequence of Brown-Zak magnetoconductance oscillations extracted from a Landau fan. Supplementary Information Figure~\ref{fig:BrownZak} shows an example, taken at $D=0.5$~V/nm, although we have confirmed the fit with several additional fans. To highlight Brown-Zak oscillations, we plot the average conductance ($\langle G_{xx} \rangle$) over the entire measured range of density versus $B$. This averaging scheme suppresses the effect of dispersing quantum Hall states. The Brown-Zak oscillations occur when $\phi/\phi_0=4B/n_s\phi_0=p/q$, where $\phi_0=h/e$ is the magnetic flux quantum and $p$ and $q$ are integers. We extract $n_s=2.55\times\mathrm{10^{12}}$~cm$^{-2}$ from a best fit of all the magnetoconductance peaks. This value is consistent with the positions of the insulating states seen on the moir\'e-adjacent side at integer moir\'e band fillings (Supplementary Information Figs.~\ref{fig:Dcomparison}c,d).

The moir\'e band filling factor $\nu$ was defined according to $\nu = n/(n_{s}/4)$. The superlattice density is related to the moir\'e wavelength by $L_M=\sqrt{8/\sqrt{3}n_s}=13.5$~nm. We then find the twist angle, $\theta=0.42^{\circ}$, from the equation \[L_M(\theta, a, \delta_{a}) = \frac{(1 + \delta_{a}) \cdot a}{\sqrt{2 \cdot (1 + \delta_{a}) \cdot \left(1 - \cos\theta\right) + \delta_a^2}}\] where $a = 0.246~\mathrm{nm}$ is the lattice constant of graphene, $\theta$ is the twist angle between the hBN and the rhombohedral graphene, and $\delta_a=\frac{a_{\mathrm{hBN}}-a}{a}\approx0.017$ is the lattice mismatch between the lattice constants of the hBN ($a_\mathrm{hBN}$) and graphene. We note that the conversion to twist angle depends on the choice of the lattice mismatch between graphene and hBN, which is not precisely known.

\textbf{Key parameters of the superconducting states.} In Extended Data Table~\ref{tab:SC_summary}, we provide simple estimates for some of the key quantities associated with superconductivity: critical temperature ($T_c$), critical out-of-plane field ($B_c$), and superconducting coherence length ($\xi$). To estimate $T_c$, we fit a line to the measured $\rho_{xx}$ in the normal state, just above the abrupt downturn. We then estimate $T_c$ as the temperature at which $\rho_{xx}$ falls to 90\% of this value. We use a similar procedure for determining $B_c$ from measurements of $\rho_{xx}$ versus $B$. We estimate the superconducting coherence length as $\xi=\sqrt{\Phi_o/(2\pi B_c)}$, where $\Phi_0$ is the superconducting magnetic flux quantum. 

To further assess the nature of superconductivity, we attempt to provide simple estimates for the mean free path in the normal state ($\ell_{mf}$) and the relevant Fermi wavevector ($k_F$). The ratio $\xi/\ell_{mf}$ parametrizes the disorder strength, where values much less than unity are generally required to support exotic order parameters. The ratio $k_F \xi$ provides insight into whether pairing is in the strong- or weak-coupling limit, with values much larger than unity signaling BCS-like (weak-coupling) behavior. We provide rough estimates of these values here, but caution that the semimetallic nature of the normal state complicates their extraction since we do not have a direct measurement of the relative densities and mobilities of the valence- and conduction-band carriers. To estimate $\ell_{mf}$, we assess the lowest magnetic field at which quantum oscillations first appear. This roughly corresponds to the condition in which the mean free path is equal to the circumference of a cyclotron orbit: $\ell_{mf} \approx 2\pi k_F \ell_B^2$, where $k_F$ is the Fermi wavevector and $\ell_B$ is the magnetic length. The Fermi wavevector can be calculated as $k_F=\sqrt{2\pi f_\nu |n|}$. 

The primarily complication is that each band has a different Fermi wavevector, and potentially a different $f_\nu$, and we do not have independent experimental measures of each of them when they coexist at the Fermi level, nor do we know for certain whether to use the value from the conduction or valence band in our estimates. As a starting point, we work under the assumption that superconductivity is driven by the conduction band and estimate $k_F$ from our calculations in R8G. The region where we observe the most robust superconductivity in R8G corresponds to $k_F \approx 0.15$~nm$^{-1}$ in the conduction band (estimated from extrapolation of the conduction-band Fermi surface from the theoretical calculation presented in Fig.~\ref{fig:2}f). Since QOs onset at $B \approx 400$~mT in Landau fans, we roughly estimate $\ell_{mf} \approx 1.5$~$\mu$m. Following this, $\xi/\ell_{mf}\sim 0.4$ and $k_F \xi\sim100$, nominally corresponding to the clean limit of weakly coupled superconductivity. 

\textbf{Superconductivity in R8G at $D<0$.} Extended Data Figure~\ref{fig:8L_negD_SC} shows zoomed-in transport measurements of the five superconducting pockets in R8G at $D<0$, which we label SC6$_8$ through SC10$_8$. Overall, they appear to be very similar to the corresponding SC1$_8$-SC5$_8$ pockets. One noteworthy difference is the region of elevated $\rho_{xx}$ at negative $n$ and $D$, which does not have an associated counterpart for $D>0$. Presently, we do not understand this feature or why it only appears for one sign of $D$. Another difference is that the SC10$_8$ pocket is much more extended than its corresponding SC5$_8$, nearly connecting with SC9$_8$. This behavior suggests that the SC9$_8$ and SC10$_8$ pockets (and, similarly, SC4$_8$ and SC5$_8$ for the other sign of $D$) are in fact one contiguous pocket, with a $T_c$ just below our base temperature in between the two.

\textbf{Transport anomalies associated with the superconducting states.} There are several transport features associated with the superconducting pockets in the R7G and R8G devices that are not anticipated for standard superconductors. First, the resistance does not reach zero in any of the pockets we study, instead saturating at a value below $100~\Omega$. As mentioned earlier, this is seen routinely in thinner crystalline graphene superconductors~\cite{Zhou2021,Patterson2025,Holleis2025,Choi2025,Zhang2025,Yang2025}. Second, for many of the pockets, the resistance above $B_c$ remains lower than at nearby gate voltages (see Supplementary Information Fig.~\ref{fig:SC3_Zoom} for analysis of SC3$_8$). In the same regime, there is additionally remnant non-linearity in d$V$/d$I$ versus $I_{dc}$ measurements. This was also seen in previous reports on thinner graphene~\cite{Holleis2025,Choi2025}. Linear transport eventually recovers at larger $B$. Third, in R8G there is an unexpected zero-bias peak in d$V$/d$I$ versus $I_{dc}$ measurements, extending far above $B_c$ and having no apparent association with superconductivity. 

We do not know the origins of these features, but speculate on some possibilities. One possible origin of the non-zero resistance is resistive heating at the contacts. There are several junctions formed between differently gated regions near the metal electrodes, and given the very low $T_c$ even modest heating could potentially result in some non-zero resistance. The zero-bias peak in the R8G device may also result from contact imperfections. As for the unusual transport features above $B_c$, they may be related to a competing many-body state that emerges upon the suppression of superconductivity. Given that superconductivity forms along a sharp resistance bump, a possibility is that this resistive state emerges above $B_c$ and also exhibits nonlinearity. Further work will be needed to better understand these observations.

\textbf{Fermiology analysis.} In order to extract information about Fermi surfaces we take fast Fourier transforms (FFTs) of $\rho_{xx}$($\frac{1}{B}$) for the Landau fans shown in Fig.~\ref{fig:1}f, and Fig.~\ref{fig:3}c, as well as Extended Data Fig.~\ref{fig:Fermiology7LVt-4V}, and Supplementary Information Fig.~\ref{fig:Fermiology7LVb2p1V}. FFTs are taken after first subtracting a fifth order polynomial from the raw longitudinal resistivity ($\rho_{xx}$) data, following the procedure established in Ref.~\onlinecite{Zhou2021_2}. The subtracted data is then interpolated onto a regular grid in order perform the FFT. For Fig.~\ref{fig:1}f the analysis is done using a magnetic field range of ($\approx0.3$~T to 1.9~T), for Fig.~\ref{fig:3}c using a range of ($\approx0.07$~T to 1~T), for Extended Data Fig.~\ref{fig:Fermiology7LVt-4V} using a range of ($\approx0.09$~T to 0.932~T), and for Supplementary Information Fig.~\ref{fig:Fermiology7LVb2p1V} using a range of ($\approx0.07$~T to 1~T) (same data as Fig.~\ref{fig:3}c). The magnetic field range for each fan is chosen to minimize noise in the FFT, without introducing any spurious frequencies. After taking the Fourier transform we normalize the raw frequencies by the Luttinger volume, $n(\frac{h}{e})$, with the normalized frequency ($f_v$) then corresponding to the fraction of the total fermi surface enclosed by a cyclotron orbit in momentum space~\cite{Zhang2023}. Consequently, the sum of all frequencies adds to one when the appropriate carrier sign, isospin degeneracies, and Fermi surface topologies are considered.

The code we use for analyzing the FFT is based on the code provided in Ref.~\onlinecite{Zhang2023}. In order to extract the relevant frequencies we further place the FFT on an interactive grid, in which individual trajectories can be fit by a user directly selecting them. We use this method to fill an array of grid coordinates which we then fit to a polynomial. A step-by-step analysis is shown in Extended Data Fig.~\ref{fig:Fermiology7LVt-4V}. In our FFT analysis we use the following color and line conventions: solid curves denote frequencies which we have directly fit using the method described above (or frequencies which directly correspond to an unchanging fraction such as $\frac{1}{4}$); dashed curves are derived from solid curves, either through multiplication, or addition/subtraction of a relevant fraction (for annular Fermi surfaces, see for example Fig.~\ref{fig:3}c). Curves (or pairs of curves) are further color coded based on the corresponding implied Fermi surface degeneracy, with red corresponding to four, blue corresponding to two, and black corresponding to other/unknown. We attempted to characterize all the prominent trajectories in each plot. However, certain regions of the fans, where the FFTs are too complicated to analyze, are left without fits. We note that interpretation of the FFTs can be ambiguous in regions of parameter space with the potential for a combination of overlapping valence and conduction bands, annular or disjoint Fermi surfaces, and isospin polarization in one or both bands.  

\textbf{Self-consistent mean-field calculation.} We present the theory for the misaligned octalayer system here and relegate the theory for the moir\'{e} heptalayer system to the Supplementary Information~\cite{Kern1999Ab,Castro2009The,2010Energy,Kindermann2010Zero,Moon2014Electronic,Jung2014Ab}. To model the band structure of $N_\ell$-layer rhombohedral graphene without a moir\'{e} structure, we adopt the conventional tight-binding parameterization wherein the Hamiltonian in the sublattice basis $\lbrace A_1, B_1, A_2, B_2,...,A_\ell,B_\ell \rbrace$ is given by \cite{Xiao2011, Min2008, Koshino2009, Guinea2006, Slizovskiy2019,mcclure1969electron} 
\begin{equation}
\begin{split}
    \mathbb{K}_{N_\ell}(\mathbf{k}) &= \begin{pmatrix}
        \mathbb{K}_1(\mathbf{k})   & \mathbb{U}_\mathrm{nl}(\mathbf{k}) & \mathbb{U}_\mathrm{nnl}  & \hdots & \mathbb{0} \\
        \mathbb{U}_\mathrm{nl}^\dagger(\mathbf{k}) & \mathbb{K}_1(\mathbf{k}) & \mathbb{U}_\mathrm{nl}(\mathbf{k}) & \hdots  & \mathbb{0}\\
        \mathbb{U}^\dagger_\mathrm{nnl}& \mathbb{U}_\mathrm{nl}^\dagger(\mathbf{k}) & \mathbb{K}_1(\mathbf{k})  & \hdots  & \mathbb{0}\\
        \vdots & \vdots  & \vdots &  \ddots  & \mathbb{0}  \\
        \mathbb{0} & \mathbb{0} & \mathbb{0}  &   \hdots & \mathbb{K}_1(\mathbf{k})
    \end{pmatrix} \\
    &+ \delta_\mathrm{dimer}\begin{pmatrix}
        \sigma_- & \mathbb{0} & \hdots & \mathbb{0} & \mathbb{0} \\
         \mathbb{0} &  \mathbb{1}& \hdots & \mathbb{0} & \mathbb{0} \\
          \vdots & \vdots &  \ddots & \vdots & \vdots  \\
         \mathbb{0} & \mathbb{0}& \hdots &  \mathbb{1} & \mathbb{0} \\
         \mathbb{0} & \mathbb{0}& \hdots & \mathbb{0} & \sigma_+
    \end{pmatrix},    
\end{split}
\end{equation}
where $\sigma_\pm$ are the Pauli lowering ($-$) and raising ($+$) operators. The  $2\times2$ matrices $\mathbb{K}_1,$ $\mathbb{U}_\mathrm{nl},$ and $\mathbb{U}_\mathrm{nnl}$ are the single-layer graphene Hamiltonian, nearest-layer hopping matrix, and next-nearest-neighbor hopping matrix respectively:
\begin{equation}
\label{eq: full tight binding}
    \begin{split}
        \mathbb{K}_1(\mathbf{k}) &= \begin{pmatrix}
        0 & -\gamma_0 f(\mathbf{k}) \\
        -\gamma_0 f^\dagger(\mathbf{k}) & 0
    \end{pmatrix}, \\
    \mathbb{U}_\mathrm{nl}(\mathbf{k}) &= \begin{pmatrix}
        -\gamma_4 f(\mathbf{k}) & -\gamma_3 f^\dagger(\mathbf{k}) \\
        \gamma_1 & -\gamma_4 f(\mathbf{k}) 
    \end{pmatrix},  \\
    \mathbb{U}_\mathrm{nnl} &= \begin{pmatrix}
        0 & \gamma_2/2 \\
        0 & 0 
    \end{pmatrix},
    \end{split}
\end{equation}
where $f(\mathbf{k}) = \sum_{i \in \lbrace 1,2,3 \rbrace} e^{i \mathbf{k} \cdot \boldsymbol{\delta}_i},$ $\boldsymbol{\delta}_1 = a/\sqrt{3}(0,1),$ $\boldsymbol{\delta}_2 = a/\sqrt{3}(-\sqrt{3}/2,-1/2),$ $\boldsymbol{\delta}_3 = a/\sqrt{3}(\sqrt{3}/2,-1/2),$ $a = 2.46$ \AA $ $ is the lattice constant, $\lbrace \gamma_0,\gamma_1,\gamma_2,\gamma_3,\gamma_4\rbrace = \lbrace 3.1,0.38,-0.015,-0.29,-0.141\rbrace$ eV are the hopping constants, and $\delta_\mathrm{dimer} = 10.5$ meV is the dimerization energy on the eclipsed sites $B_1,A_2,B_2,..., A_{N_\ell-1}, B_{N_\ell-1},A_{N_\ell}$ (only $A_1$ and $B_{N_\ell}$ are not eclipsed) \cite{Zhou2021_2}.  The layer-dependent on-site potential energy due to a vertical displacement field is added to the Hamiltonian using the following potential
\begin{equation}
    \mathbb{\Delta} = \frac{\Delta}{2} \begin{pmatrix}
        +(N_\ell-1) \mathbb{1} & \mathbb{0} & \hdots & \mathbb{0} \\
         \mathbb{0} & +(N_\ell-3)  \mathbb{1} & \hdots  & \mathbb{0} \\
           \vdots & \vdots &  \ddots & \vdots   \\
         \mathbb{0} & \mathbb{0} &\hdots  & -(N_\ell-1)  \mathbb{1}
    \end{pmatrix},
\end{equation}
where $\Delta$ is the energy difference between two consecutive layers. The energy difference across the entire $N_\ell$-layer stack is $\Delta (N_\ell-1).$

To account for band renormalization due to electron-electron interactions, we implement self-consistent mean-field theory within the Hartree-Fock approximation. The mean-field Hamiltonian in second-quantized notation is 
\begin{equation}
\begin{split}
\label{eq: Hartree Fock Hamiltonian}
    \hat{\mathcal{H}}_\mathrm{HF} = \sum_{\substack{\mathbf{k} , \alpha,\ell, \alpha',\ell' }}  &\hat{c}_{\alpha,\ell,\mathbf{k}}^\dagger  \left[ \mathbb{K}^{\alpha,\ell, \alpha',\ell'}(\mathbf{k}) \right. \\
    &+ \left. \mathbb{H}^{\alpha,\ell, \alpha',\ell'} + \mathbb{F}^{\alpha,\ell, \alpha',\ell'}(\mathbf{k})  \right] \hat{c}_{\alpha',\ell',\mathbf{k}}.    
\end{split}
\end{equation}
Here, $\hat{c}^\dagger_{\alpha,\ell,\mathbf{k}}$ is the creation operator for a plane-wave state with $\alpha$ quantum numbers, where $\alpha$ is a multi-index label that includes valley $\xi$, spin $s$, and sublattice $\sigma$, on layer $\ell$ and with momentum $\mathbf{k}.$ $\mathbb{K}^{\alpha,\ell, \alpha',\ell'}(\mathbf{k})$ is the same as $\mathbb{K}_{N_\ell}(\mathbf{k})$ with internal indices explicitly specified. The Hartree and Fock matrices are given by 
\begin{equation}
\label{eq: Hartree and Fock terms}
    \begin{split}
        \mathbb{H}^{\alpha,\ell, \alpha',\ell'}  &= +\frac{\delta_{\alpha, \alpha'}\delta_{\ell,\ell'}}{V_\mathrm{sys}} \sum_{\mathbf{p} ,  \beta,\ell''}\mathcal{V}_{\ell,\ell''}(\mathbf{0}) \mathbb{D}^{\beta ,\ell'',\beta,\ell''}(\mathbf{p}),\\
        \mathbb{F}^{\alpha,\ell, \alpha',\ell'}(\mathbf{k}) &= - \frac{1}{V_\mathrm{sys}} \sum_{\mathbf{p}}\mathcal{V}_{\ell,\ell'}\left(\mathbf{p}  - \mathbf{k} \right) \mathbb{D}^{\alpha,\ell, \alpha',\ell'}(\mathbf{p}),
    \end{split}
\end{equation}
where $V_\text{sys}$ is the volume of the system. We have defined a density matrix $\mathbb{D}^{\alpha,\ell ,\alpha ',\ell'}(\mathbf{k}) = \langle \hat{c}^\dagger_{\alpha',\ell',\mathbf{k}} \hat{c}_{\alpha, \ell, \mathbf{k}} \rangle;$ we draw attention to the order of the indices in our definition of the density matrix. Explicitly, the matrix elements can be calculated from the coefficients of the one-particle eigenstates $\mathbb{D}^{\alpha,\ell, \alpha',\ell'}(\mathbf{k}) = \sum_{n \in \mathrm{occupied}}  \phi_{\alpha, \ell,\mathbf{k} }^{(n)}\phi_{\alpha',\ell', \mathbf{k}}^{(n)*},$ where the eigenstates written in terms of plane-wave basis functions $\ket{\psi_{\alpha,\ell,\mathbf{k}} }$ are given by  $\ket{\psi_{ \mathbf{k}}^{(n)}} = \sum_{\alpha,\ell} \phi^{(n)}_{\alpha,\ell, \mathbf{k}} \ket{\psi_{\alpha,\ell,\mathbf{k}} }$ and $n$ labels band index. Normalization requires $\sum_{\alpha,\ell}\left|\phi_{\alpha,\ell,\mathbf{k}}^{(n)}\right|^2 = 1.$ To avoid over-counting the effects of Coulomb interactions already included in \textit{ab initio} calculations, we subtract from the density matrices in Eq. \eqref{eq: Hartree and Fock terms} a uniform background density: $\mathbb{D} \mapsto \mathbb{D}- \mathbb{D}_\text{ref},$ where $\mathbb{D}_\text{ref}^{\alpha,\ell, \alpha',\ell'}(\mathbf{p}) = \frac{1}{2} \delta_{\ell,\ell'} \delta_{\alpha,\alpha'}$ \cite{Xie2020Nature,Bernevig2021Twisted}. 

In a multilayer system, especially with a large number of layers, charges can reorganize across the layers to oppose the externally applied displacement field. To model this effect, we use a layer-dependent Coulomb potential (e.g. see Ref. \citenum{phongC2} for derivation)
\begin{equation}
\begin{split}
 \label{eq: Coulomb potential}
    \mathcal{V}_{\ell,\ell'}(\mathbf{q}) = \frac{e^2\mathrm{csch} \left( |\mathbf{q}| d_g\right)}{2 \epsilon_0 \epsilon_r|\mathbf{q}|}   &\left[ \cosh \left(|\mathbf{q}|\{d_g-|z_\ell-z_{\ell'}|\}\right) \right.\\
    &\left.- \cosh \left(|\mathbf{q}|\{d_g-|z_\ell+z_{\ell'}|\}\right)  \right],   
\end{split}
\end{equation}
where $e$ is the electric charge, $\epsilon_0 \epsilon_r$ is the dielectric constant, $d_g$ is the distance between the top and bottom gates, and $z_\ell$ is the vertical height of the $\ell$ layer measured from the bottom gate situated at $z = 0$. The $\mathbf{k} = 0$ component of this potential,
\begin{equation}
 \mathcal{V}_{\ell,\ell'}(\mathbf{0}) = \lim_{\mathbf{q} \rightarrow \mathbf{0}}\mathcal{V}_{\ell,\ell'}(\mathbf{q}) = \frac{e^2}{\epsilon_0\epsilon_r} \left[\min\left(z_\ell,z_{\ell'} \right)-\frac{z_\ell z_{\ell'}}{d_g} \right],
\end{equation}
reproduces the screening of uniform charges in a classical capacitor model \cite{Kolar2023}. In this work, we use $d_g = 40$ nm and assume that the graphene stack is located symmetrically between the two gates. The dielectric constant is chosen to be  $\epsilon_r = 16$. This value is chosen to reproduce the onset of insulating behavior at charge neutrality seen in the experiment. Since our calculation does not directly compute internal screening due to both active (involved in HF) and remote bands, this higher value of $\epsilon_r$ compared to the known values of hBN's dielectric constant allows us to qualitatively account for screening not only  from the dielectric environment but also between the electrons themselves \cite{parra2025band}.  

In numerical simulation, we determine Eq. \eqref{eq: Hartree Fock Hamiltonian} iteratively. We impose an ultraviolet cutoff by choosing a regular hexagonal region in reciprocal space centered at each valley $K_\xi = \xi (4\pi/3a,0)$ with sides equal to $0.1\times|K_\xi| = 0.17$ \AA$^{-1}.$ Inside each such region, $3N_{\mathbf{k}}(N_{\mathbf{k}}+1)+1$ points are sampled in the calculation. We choose $N_{\mathbf{k}} = 70,$ which means that our grid contains $14,911$ points per valley and spin. In this work, we do not search for symmetry-broken states; therefore, our ansatz is always the symmetric state where all isospins are filled equally. To facilitate faster convergence, at each step, we use a weighted average of the most recent two density matrices. For the calculation of the layer-resolved DOS and the total DOS, we use the following approximations
\begin{subequations}
\begin{align}
     \mathrm{DOS}_\ell(\varepsilon) &= \frac{1}{\pi} \sum_{n, \mathbf{k}} \left[\sum_\alpha|\phi^{(n)}_{\alpha,\ell,\mathbf{k}}|^2 \right] \frac{\eta}{\eta^2+(\varepsilon-\varepsilon_{n,\mathbf{k}})^2}, \\
     \mathrm{DOS}(\varepsilon) &= \sum_\ell\mathrm{DOS}_\ell(\varepsilon),
\end{align}
\end{subequations}
where $\eta$ is a broadening factor chosen to be $1-2$ meV. These definitions have units of inverse energy; there is no factor of $1/V_\mathrm{sys}$ since we have integrated over the lateral directions. The small broadening factor is necessary to produce a smooth evolution of the DOS in the phase diagram since our simulation mesh is necessarily granular (though it is quite large as aforementioned). 

\textbf{Theoretical determination of displacement field.} The experimental phase diagrams are plotted as a function of the displacement field $D$, which is related to the top and bottom gate charge densities, $\sigma_t$ and $\sigma_b,$ that are extracted experimentally using the gate capacitances:
\begin{equation}
\label{eq: displacement field}
    D = \frac{\sigma_t-\sigma_b}{2\epsilon_0}.
\end{equation}
In our theoretical calculations, $D$ is not a direct input, but is instead only represented by its proxy $\Delta$. To translate between the two quantities so that theoretical calculations can be compared to experimental data, we use the Hartree approximation wherein we assume that $D$ is determined only by the average uniform charge density on each layer, i.e. charge variation on the unit-cell scale does not matter \cite{McCann2006Asymmetry,Gava2009Ab}. Let us denote $D_{\ell, \ell'}$ the displacement field between layers $\ell$ and $\ell'.$ By Gauss's Law, we therefore have
\begin{equation}
\label{eq: Gauss's Law}
    \frac{\sigma_\ell}{\epsilon_0} = D_{\ell,\ell+1} - D_{\ell-1,\ell},
\end{equation}
where $\sigma_\ell$ is the charge density on layer $\ell.$ Here, $\ell \in \lbrace 1,...,N_\ell\rbrace$ denotes graphene layers, and $\ell = 0$ is the top gate and $\ell = N_\ell+1$ is the bottom gate. Charge neutrality enforces $\sum_{\ell=0}^{N_\ell+1}\sigma_\ell = 0.$ Substituting Eq. \eqref{eq: Gauss's Law} into Eq. \eqref{eq: displacement field}, we find
\begin{equation}
\label{eq: D field in terms of D_t and D_b}
    D = \frac{D_\text{t} + D_\text{b}}{2},
\end{equation}
where $D_\text{t} = D_{0,1}$ and $D_\text{b} = D_{N_\ell,N_\ell+1}.$ We can relate these two fields closest to the gates to any other two fields between the graphene layers. For instance, we can rewrite Eq. \eqref{eq: D field in terms of D_t and D_b} as 
\begin{equation}
    D = \frac{\sigma_{N_\ell}-\sigma_1}{2\epsilon_0} + \frac{D_{N_\ell-1,N_\ell}+D_{1,2}}{2};
\end{equation}
more generally, for any $n,n'\in \lbrace 1,...,N_\ell\rbrace,$ we have
\begin{equation}
\label{eq: recursive formula}
    D = \frac{\sum_{\ell = n'}^{N_\ell}\sigma_{\ell}-\sum_{\ell=1}^n\sigma_\ell}{2\epsilon_0} + \frac{D_{n'-1,n'}+D_{n,n+1}}{2}.
\end{equation}
The layer charge densities can be calculated by tracing over the converged density matrices. The interlayer displacement fields are calculable from the onsite energies $\varepsilon_\ell = \mathbb{H}^{\alpha,\ell,\alpha,\ell} +\mathbb{\Delta}^{\alpha,\ell,\alpha,\ell}:$
\begin{equation}
\label{eq: electric potential to displacement field}
    D_{\ell,\ell+1} = \frac{\epsilon_r}{ed}\left( \varepsilon_{\ell+1} - \epsilon_\ell\right),
\end{equation}
where $d = 3.35$ \AA $\text{ }$ is the distance between graphene layers. Together, Eqs. \eqref{eq: recursive formula} and \eqref{eq: electric potential to displacement field} allow us to calculate $D$ for every $\Delta$ using the self-consistent mean-field solution at said $\Delta.$ We emphasize that, in our approach, once the hopping parameters and dielectric constant, which controls the strength of Coulomb interactions at zero and finite $\mathbf{q},$ are specified (and these are constrained by experimental considerations), $D$ and $\Delta$ are unambiguously related to each other by self-consistency. 

\section*{Acknowledgments}

The authors thank UBC research associate Silvia Folk for helpful contributions, as well as Roger Waleffe for useful discussions about the FFT analysis code. Research on superconductivity was supported by the Army Research Office under award number W911NF-25-1-0012. Work on topology was supported as part of Programmable Quantum Materials, an Energy Frontier Research Center funded by the U.S. Department of Energy (DOE), Office of Science, Basic Energy Sciences (BES), under award DE-SC0019443. Device fabrication was supported by National Science Foundation (NSF) CAREER award no. DMR-2041972 and University of Washington Molecular Engineering Materials Center, a U.S. National Science Foundation Materials Research Science and Engineering Center (DMR-2308979). Experiments at the University of British Columbia were undertaken with support from the Natural Sciences and Engineering Research Council of Canada; the Canada Foundation for Innovation; the Canadian Institute for Advanced Research; the Max Planck-UBC-UTokyo Centre for Quantum Materials and the Canada First Research Excellence Fund, Quantum Materials and Future Technologies Program; and the European Research Council (ERC) under the European Union’s Horizon 2020 research and innovation program, Grant Agreement No. 951541. A.O. and M.Y. acknowledge support from the State of Washington-funded Clean Energy Institute. Numerical calculations are done using the High Performance Compute Cluster of the Research Computing Center (RCC) at Florida State University. V.T.P. and C.L. are supported by start-up funds from Florida State University and the National High Magnetic Field Laboratory. The National High Magnetic Field Laboratory is supported by the National Science Foundation through NSF/DMR-2128556 and the State of Florida. K.W. and T.T. acknowledge support from the JSPS KAKENHI (Grant Numbers 21H05233 and 23H02052) and World Premier International Research Center Initiative (WPI), MEXT, Japan. This work made use of shared fabrication facilities at UW provided by NSF MRSEC 2308979.

\section{Author Contributions} 
M.K. fabricated the samples with assistance from A.O. M.K., D.W. and A.O. led measurements at UBC remotely with R.T. assisting on-site. V.T.P. performed the theoretical calculations supervised by C.L. K.W. and T.T. provided the hBN crystals. J.F. and M.Y. supervised the project. 

\section*{Competing interests}
The authors declare no competing interests.

\section*{Additional Information}
Correspondence and requests for materials should be addressed to Joshua Folk or Matthew Yankowitz.

\section*{Data Availability}
Source data are available for this paper. All other data that support the findings of this study are available from the corresponding author upon request.

\bibliographystyle{naturemag}
\bibliography{references}

\newpage

\renewcommand{\figurename}{Extended Data Fig.}
\renewcommand{\thesubsection}{S\arabic{subsection}}
\setcounter{secnumdepth}{2}
\renewcommand{\thetable}{S\arabic{table}}
%\subsubsectionfont{\normalfont\large\itshape\underline}
%\renewcommand{\theequation}{S\arabic{equation}}
\setcounter{figure}{0} 
\setcounter{equation}{0}

\onecolumngrid
\newpage

\FloatBarrier           
\section*{Extended Data}
\FloatBarrier     

%%%%%%%%%%%%%%%%%%%%%%%%%%%%%%%%%%%%%%%%%%%%%%%%%%%%%%%%%%%%%%%%%%%%%%
\begin{table*}[h]
\centering
\setlength{\tabcolsep}{20pt}
\begin{tabular}{cccc} 
\hline\hline
\rule{0pt}{2.8ex}%
pocket & $T_c$ (mK) & $B_c$ (mT) & $\xi$ (nm) \\[0.6ex]
\hline
\rule{0pt}{2.8ex}%
~~
SC1$_7$ & 89 & 1.5 & 470  \\
~~
SC2$_7$ & 51 & 2.0 & 410  \\
~~
SC1$_8$ & 40 & 0.65 & 710  \\
~~
SC2$_8$ & 67 & 1.0 & 570  \\
~~
SC3$_8$ & 170 & 0.8 & 640  \\
~~
SC4$_8$ & 20 & 0.15 & 1480  \\
~~
SC5$_8$ & 78 & 0.32 & 1010  \\[0.6ex]
\hline\hline
\end{tabular}
\caption{\textbf{Summary of key parameters relevant to superconductivity.}
Values are extracted following procedures described in the Methods.
}
\label{tab:SC_summary}
\end{table*}
%%%%%%%%%%%%%%%%%%%%%%%%%%%%%%%%%%%%%%%%%%%%%%%%%%%%%%%%%%%%%%%%%%%%%%

\begin{figure*}[h]
\includegraphics[width=0.8\textwidth]{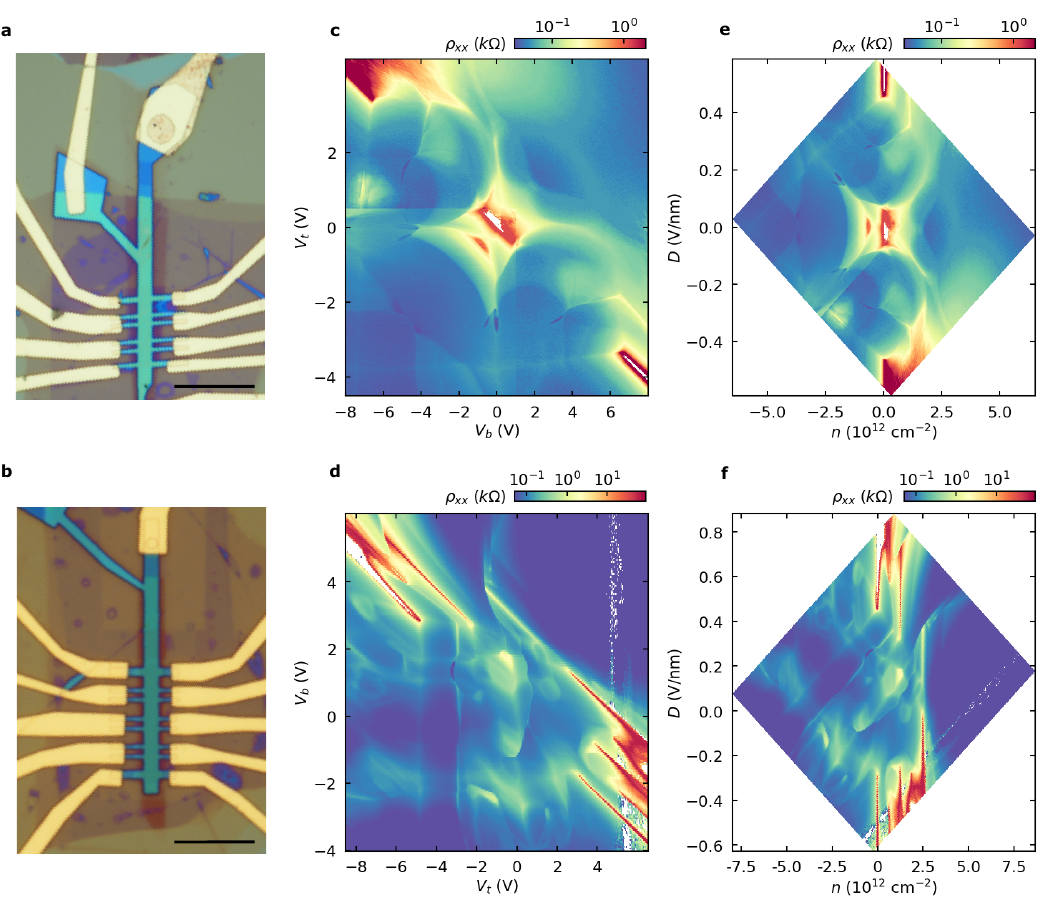} 
\caption{\textbf{Optical micrographs and full gate maps of both devices.}
\textbf{a}, Optical micrograph of the misaligned R8G device. 
\textbf{b}, Same as \textbf{(a)} for the moir\'e R7G device. For both, the scale bar is 10~$\mu$m. 
\textbf{c}, Map of $\rho_{xx}$ for the R8G device measured over a wide range of $V_b$ and $V_t$. The map is acquired in pieces and stitched together, with different silicon gate voltages in order to optimally reduce the contact resistance in each quadrant.
\textbf{d}, Same as \textbf{(c)} for the R7G device.
\textbf{e}, The map from \textbf{(c)} converted to $n-D$ axes (see Methods for conversion).
\textbf{f}, Same as \textbf{(e)} for the R7G device.
}
\label{fig:ggmap7L_8L}
\end{figure*}

\begin{figure*}[t]
\includegraphics[width=\textwidth]{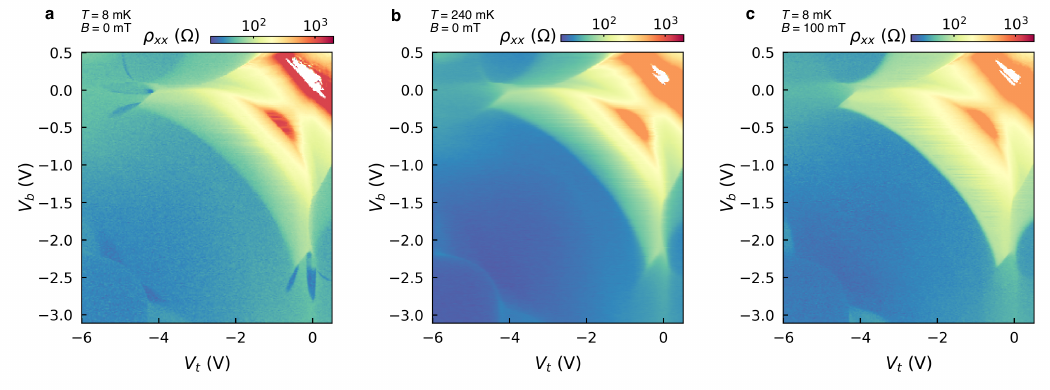} 
\caption{\textbf{Transport with varying temperature and magnetic field in the R8G device.}
\textbf{a}, Map of $\rho_{xx}$ versus $V_b$ and $V_t$ at base temperature and $B=0$. The highly resistive feature in the top right is the correlated insulator at the CNP surrounding $D=0$. Superconducting pockets SC1$_8$ to SC4$_8$, and their corresponding $D<0$ counterparts, are visible within this region of gate voltage.
\textbf{b}, The same map as \textbf{(a)} at $T=240$~mK, above $T_c$ of all the pockets.
\textbf{c}, The same map as \textbf{(a)} at $B=100$~mT, above $B_c$ of all the pockets.
}
\label{fig:T_B_comparison}
\end{figure*}

\begin{figure*}[t]
\includegraphics[width=\textwidth]{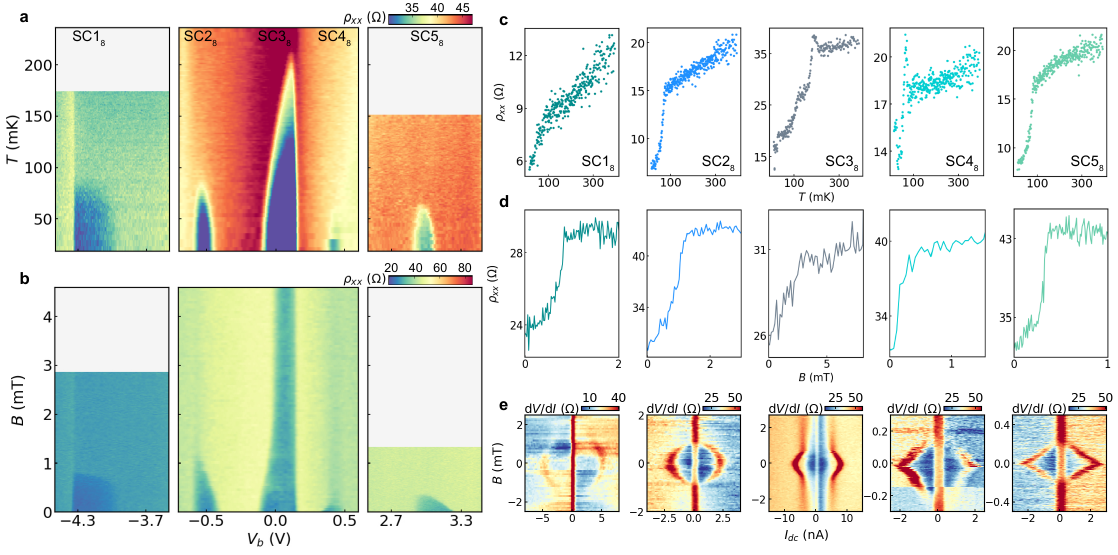} 
\caption{\textbf{Additional characterization of SC1$_8$ to SC5$_8$ superconducting pockets in the R8G device.}
\textbf{a}, Maps of $\rho_{xx}$ versus $V_b$ and $T$ around each pocket, acquired with fixed $V_t=-2.75$~V for SC1$_8$ and $V_t=-2.50$~V for SC2$_8$-SC5$_8$.
\textbf{b}, The same maps versus $B$. 
\textbf{c}, Measurement of $\rho_{xx}$ versus $T$ at optimal doping of each pocket. The measurements are acquired with a small $I_{dc}<1$~nA to mitigate the effects of a zero-bias anomaly. 
\textbf{d}, The same measurements versus $B$, with $I_{dc}=0$.
\textbf{e}, Maps of d$V$/d$I$ versus $I_{dc}$ and $B$ for each pocket. In addition to sharp resistive features forming diamond-like structures arising due to superconductivity, there is also an unexpected anomaly at $I_{dc}=0$.
}
\label{fig:8L_allSC)}
\end{figure*}

\begin{figure*}[t]
\includegraphics[width=0.95\textwidth]{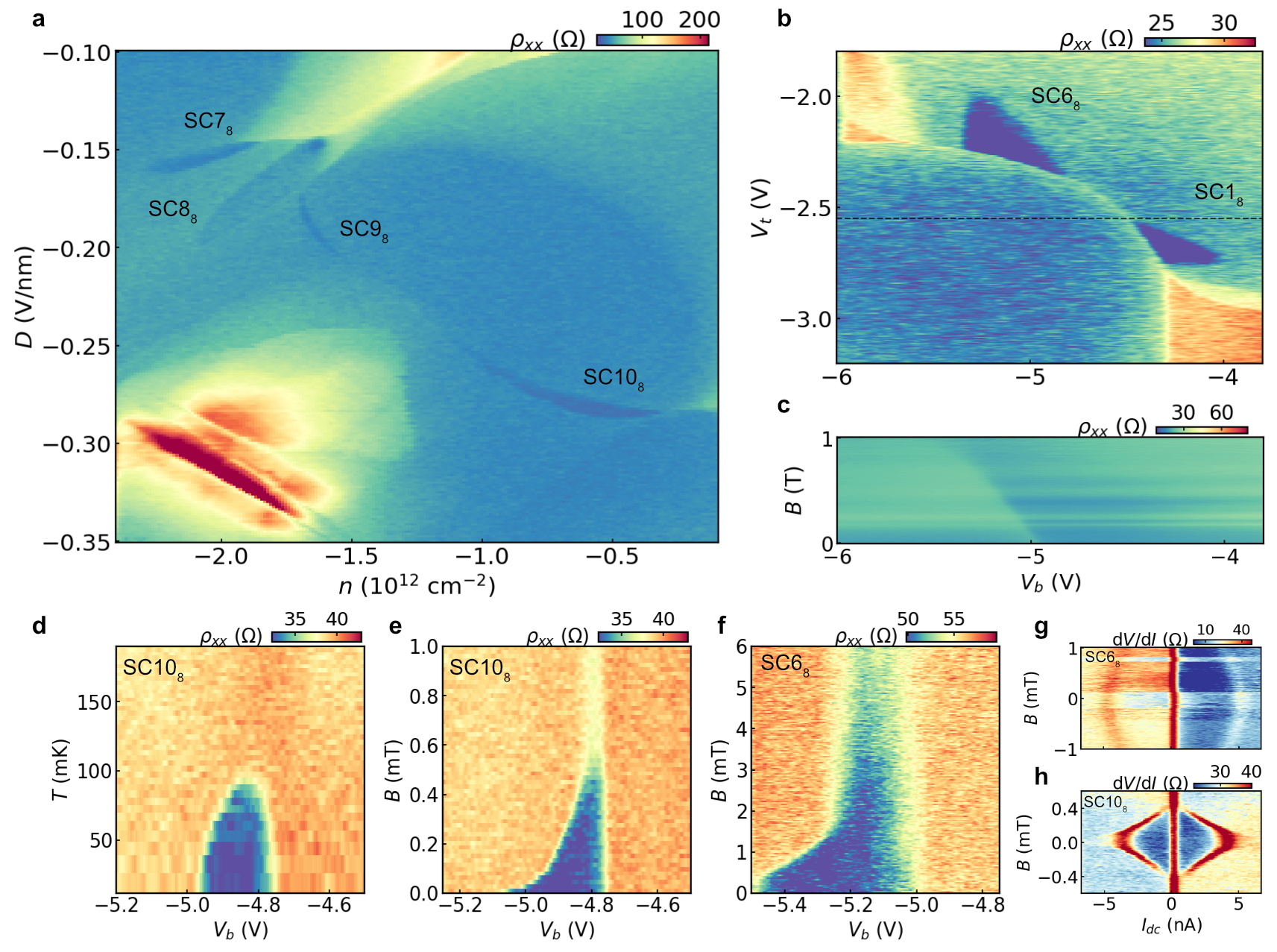} 
\caption{\textbf{Additional characterization of superconducting pockets at $D<0$ in the R8G device.}
\textbf{a}, Map of $\rho_{xx}$ versus $n$ and $D$ corresponding to $D<0$. The map contains four superconducting pockets, SC7$_8$ to SC10$_8$, which correspond to SC2$_8$ to SC5$_8$ for $D>0$. The feature in the bottom left corner exhibits unusual nonlinear behavior and does not appear for $D>0$; at present, we do not know its origin.
\textbf{b}, Map of $\rho_{xx}$ versus $V_b$ and $V_t$ showing SC1$_8$ and SC6$_8$.
\textbf{c}, Landau fan diagram taken by sweeping $V_b$ with fixed $V_t=-2.55$~V, corresponding to the black dashed line in \textbf{(b)}. The sharp resistance step associated with the edge of SC1$_8$ slopes to the left with $B$, suggestive of a spin-polarized half-metal to the right and an isospin unpolarized phase to the left.
\textbf{d}, Measurement of $\rho_{xx}$ versus $V_b$ and $T$ at fixed $V_t=1.6$~V and $B=0$ for SC10$_8$.
\textbf{e}, Similar measurement versus $B$ at base temperature.
\textbf{f}, Similar measurement as \textbf{(d)} for the SC6$_8$ pocket at fixed $V_t=-2.2$~V.
\textbf{g}, Measurement of d$V$/d$I$ versus $I_{dc}$ and $B$ for SC6$_8$.
\textbf{h}, Similar measurement for SC10$_8$. Both exhibit the similar zero bias anomalies as seen for the $D>0$ pockets.
}
\label{fig:8L_negD_SC}
\end{figure*}

\begin{figure*}[t]
\includegraphics[width=\textwidth]{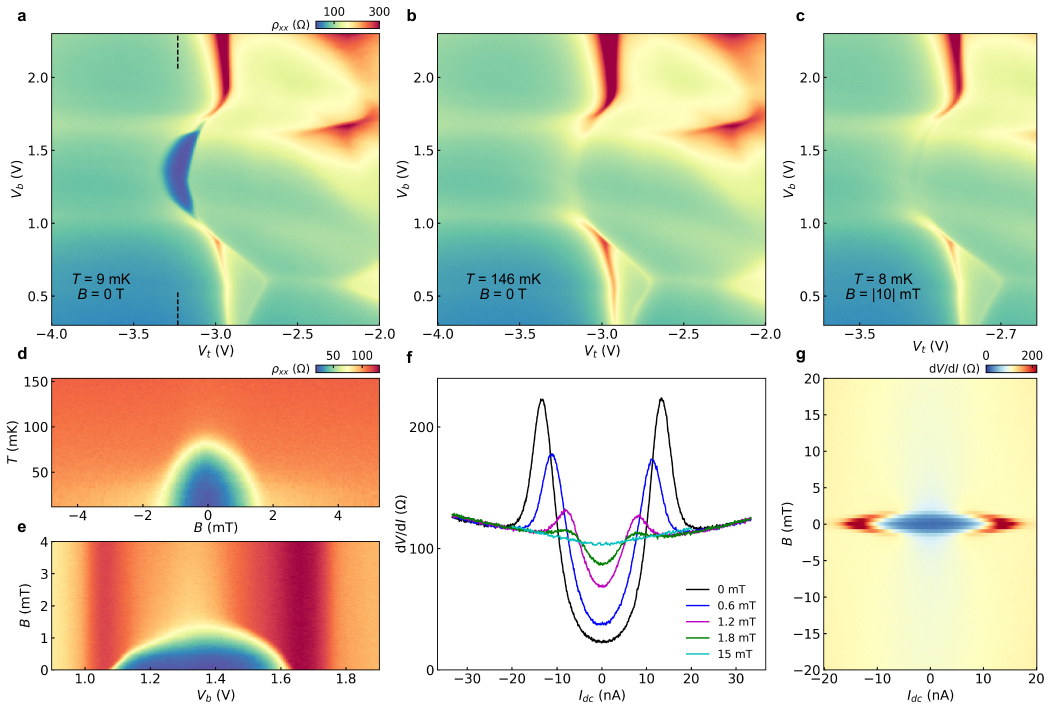} 
\caption{\textbf{Additional characterization of the SC1$_7$ pocket from the moir\'e R7G device.}
\textbf{a}, Map of $\rho_{xx}$ versus $V_b$ and $V_t$ at base temperature and $B=0$.
\textbf{b}, Similar measurement at $T=146$~mK, above $T_c$ of SC1$_7$.
\textbf{c}, Similar measurement, over a narrowed range of $V_t$, at base temperature and symmetrized at $|B|=10$~mT, above $B_c$.
\textbf{d}, Measurement of $\rho_{xx}$ versus $B$ and $T$ at optimal doping of SC1$_7$ ($V_b=1.40$~V and $V_t=-3.22$~V).
\textbf{e}, Measurement of $\rho_{xx}$ versus $V_b$ and $B$ at base temperature and fixed $V_t=-3.23$~V, corresponding to the line trace indicated by black dashed lines in \textbf{(a)}. In this map, SC1$_7$ is bounded by two resistance bumps associated with nearly constant $V_b$, i.e., the surface associated with the valence band.
\textbf{f}, Measurements of d$V$/d$I$ versus $I_{dc}$ at selected values of $B$ taken at optimal doping of SC1$_7$ ($V_b=1.40$~V and $V_t=-3.22$~V). 
\textbf{g}, Map of d$V$/d$I$ versus $I_{dc}$ and $B$.
}
\label{fig:SIalignedSC}
\end{figure*}

\begin{figure*}[t]
\includegraphics[width=\textwidth]{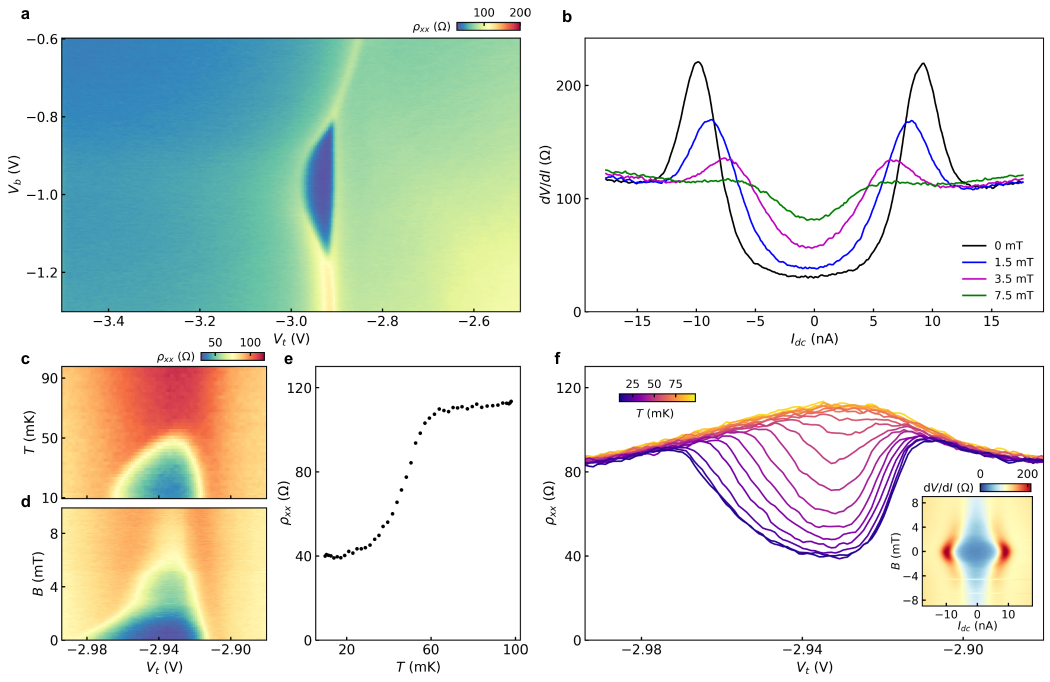} 
\caption{\textbf{Additional characterization of the SC2$_7$ pocket from the moir\'e R7G device.}
\textbf{a}, Map of $\rho_{xx}$ versus $V_b$ and $V_t$.
\textbf{b}, Measurements of d$V$/d$I$ versus $I_{dc}$ for several selected values of $B$. 
\textbf{c}, Measurement of $\rho_{xx}$ versus $V_t$ and $T$ at fixed $V_b=-0.95$~V and $B=0$.
\textbf{d}, Similar measurement as \textbf{(c)} versus $B$ at base temperature.
\textbf{e}, Measurement of $\rho_{xx}$ versus $T$ (taken at $V_b=-0.95$~V and $V_t=-2.93$~V).
\textbf{f}, Measurement of $\rho_{xx}$ versus $V_t$ at several selected values of $T$. (Inset) Map of d$V$/d$I$ versus $I_{dc}$ and $B$ (taken at $V_b=-0.95$~V and $V_t=-2.93$~V). 
}
\label{fig:SC7_2}
\end{figure*}

\begin{figure*}[h]
\includegraphics[width=\textwidth]{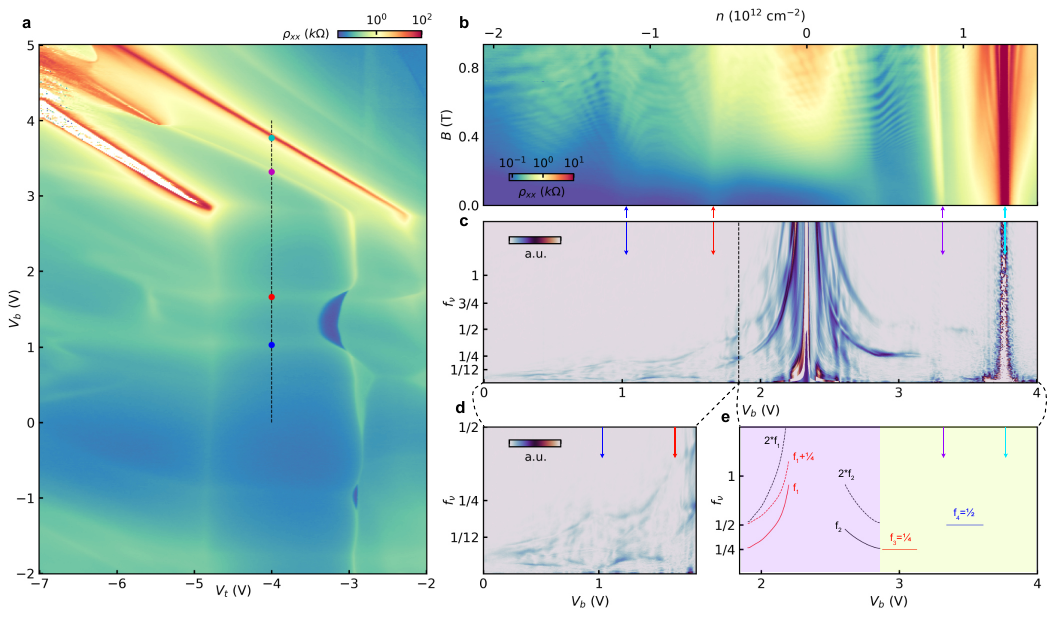} 
\caption{\textbf{Additional Fermiology analysis for the moir\'e R7G device.}
\textbf{a}, Map of $\rho_{xx}$ versus $V_b$ and $V_t$ taken at base temperature and $B=0$. 
\textbf{b}, Landau fan taken by sweeping $V_b$ at fixed $V_t=-4$~V, corresponding to the black dashed line in \textbf{(a)}. Red, blue, purple, and cyan arrows denote the positions of key features, corresponding to the red, blue, purple, and cyan dots from \textbf{(a)}.
\textbf{c}, FFT of the Landau fan.
\textbf{d}, Zoom-in of the FFT between $V_b=0$ and $1.84$~V. The dominant frequencies in the FFT abruptly change between the red and blue arrows, indicating a Fermi surface reconstruction in the region where SC1$_7$ forms at nearby $V_t$.
\textbf{e}, Schematic of the dominant frequencies from the FFT between $V_b=1.84$~V and $4.00$~V. For $V_b>2.8$~V, corresponding to crossing the jet, the dominant frequency is $0.25$ indicating a single, simply-connected Fermi surface from the conduction band. For $V_b>3.32$~V, upon crossing the sharp resistive bump feature associated with superconductivity and the IQAH state, the dominant frequency shifts to $0.5$, indicating a half-metal state. The associated feature drifts to the left in the Landau fan, consistent with a spin-polarized half-metal state.
}
\label{fig:Fermiology7LVt-4V}
\end{figure*}

\begin{figure*}[t]
\includegraphics[width=0.9\textwidth]{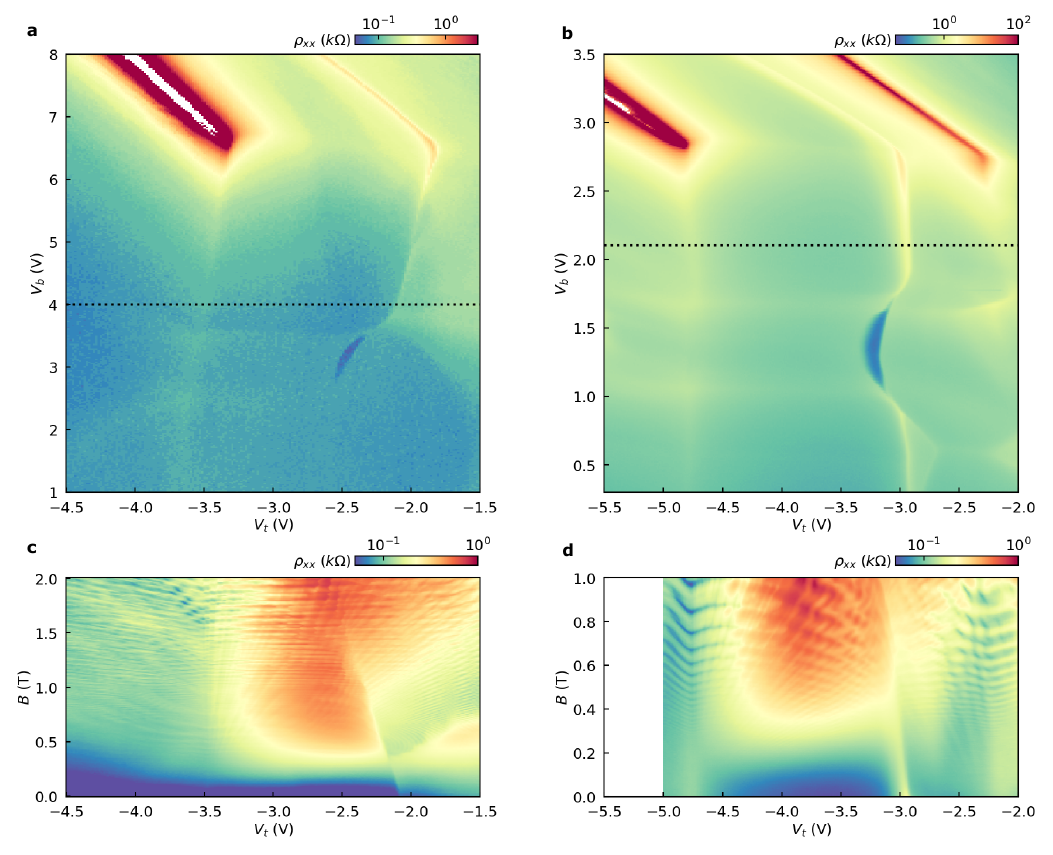} 
\caption{\textbf{Comparison of SC5$_8$ from the R8G device and SC1$_7$ from the moir\'e R7G device.}
\textbf{a}, Map of $\rho_{xx}$ versus $V_b$ and $V_t$ around SC5$_8$ from the R8G device. 
\textbf{b}, Similar measurement around SC1$_7$ from the R7G device. In both, the diagonal insulating feature in the top left corresponds to the band insulator at the CNP. The resistive feature to its left corresponds is diagonal at large $V_b$, but becomes nearly vertical at smaller $V_b$ in the band-overlapping regime. In both cases, this feature becomes superconducting when it is intersected by a horizontal resistive bump. the pocket of superconductivity closes when it intersects with a second horizontal resistive bump at smaller $V_b$. The similarities of these two pockets suggests they share similar origins.
\textbf{c}, Landau fan taken by sweeping $V_t$ at fixed $V_b=4.00$~V, along the black dashed line in \textbf{(a)}.
\textbf{d}, Landau fan taken by sweeping $V_t$ at fixed $V_b=2.10$~V, along the black dashed line in \textbf{(b)}. In both, the sharp resistive bump feature that develops into superconductivity drifts to the left with $B$, consistent with a spin-polarized half-metal to the right gaining Zeeman energy over an unpolarized phase to the left.
}
\label{fig:R7G_R8G_SC_compare}
\end{figure*}

\begin{figure*}[t]
\includegraphics[width=\textwidth]{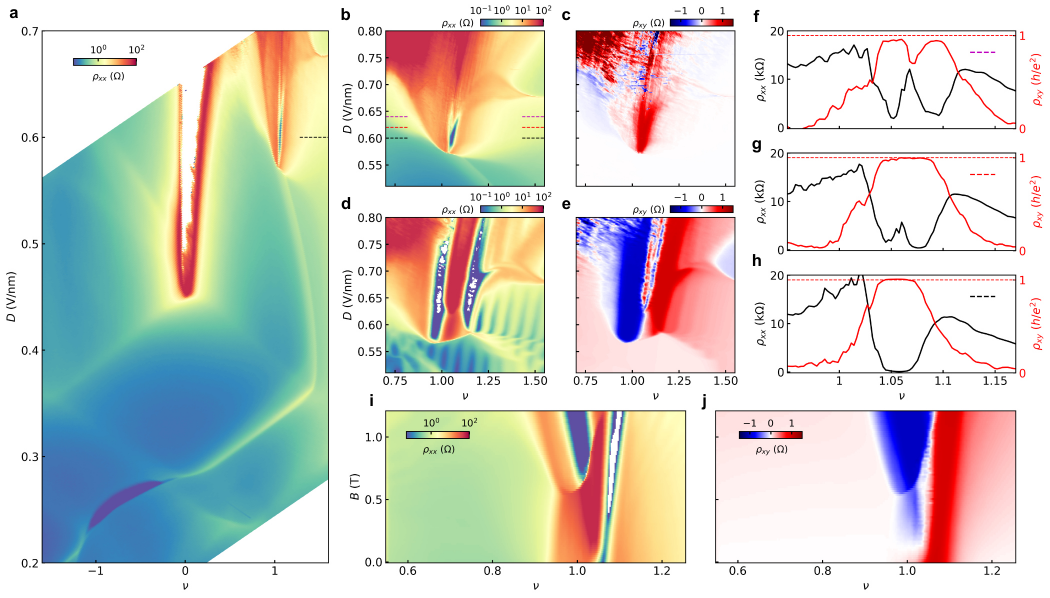} 
\caption{\textbf{Additional analysis of the IQAH state in the moir\'e R7G device.}
\textbf{a}, Map of $\rho_{xx}$ versus $\nu$ and $D$ taken at base temperature and $B=0$.
\textbf{b}, Zoom-in of the region around $\nu=1$ at large $D$ ($B=0$).
\textbf{c}, Map of $\rho_{xy}$ over the same region as \textbf{(b)} ($B=0$).
\textbf{d}, Same as \textbf{(b)} symmetrized at $B=2$~T.
\textbf{e}, Same as \textbf{(c)} antisymmetrized at $B=2$~T.
\textbf{f}, Measurements of $\rho_{xx}$ and $\rho_{xy}$ versus $\nu$ taken at $B=0$ at the positions of the corresponding color-coded dashed lines in \textbf{(b)}. At $D=0.60$~V/nm there is a single contiguous region near $\nu=1.05$ with $\rho_{xy}\approx h/e^2$ and small $\rho_{xx}$. At larger $D$, the state splits into two, with a resistive bump separating them. 
\textbf{i}, Landau fan of $\rho_{xx}$ taken at $D=0.6$~V/nm, corresponding do the black dashed line in \textbf{(a)}. 
\textbf{j}, The same measurement as \textbf{(i)}, but of $\rho_{xy}$. The fans show the $C=+1$ state arising from $B=0$, and the abrupt emergence of a $C=-1$ state above $B\approx0.6$~T. The coexistence of these two states at modest $B$ can also be seen from the maps in \textbf{(d)}-\textbf{(e)}.
}
\label{fig:7L_IQAH}
\end{figure*}

\begin{figure*}[t]
\includegraphics[width=0.95\textwidth]{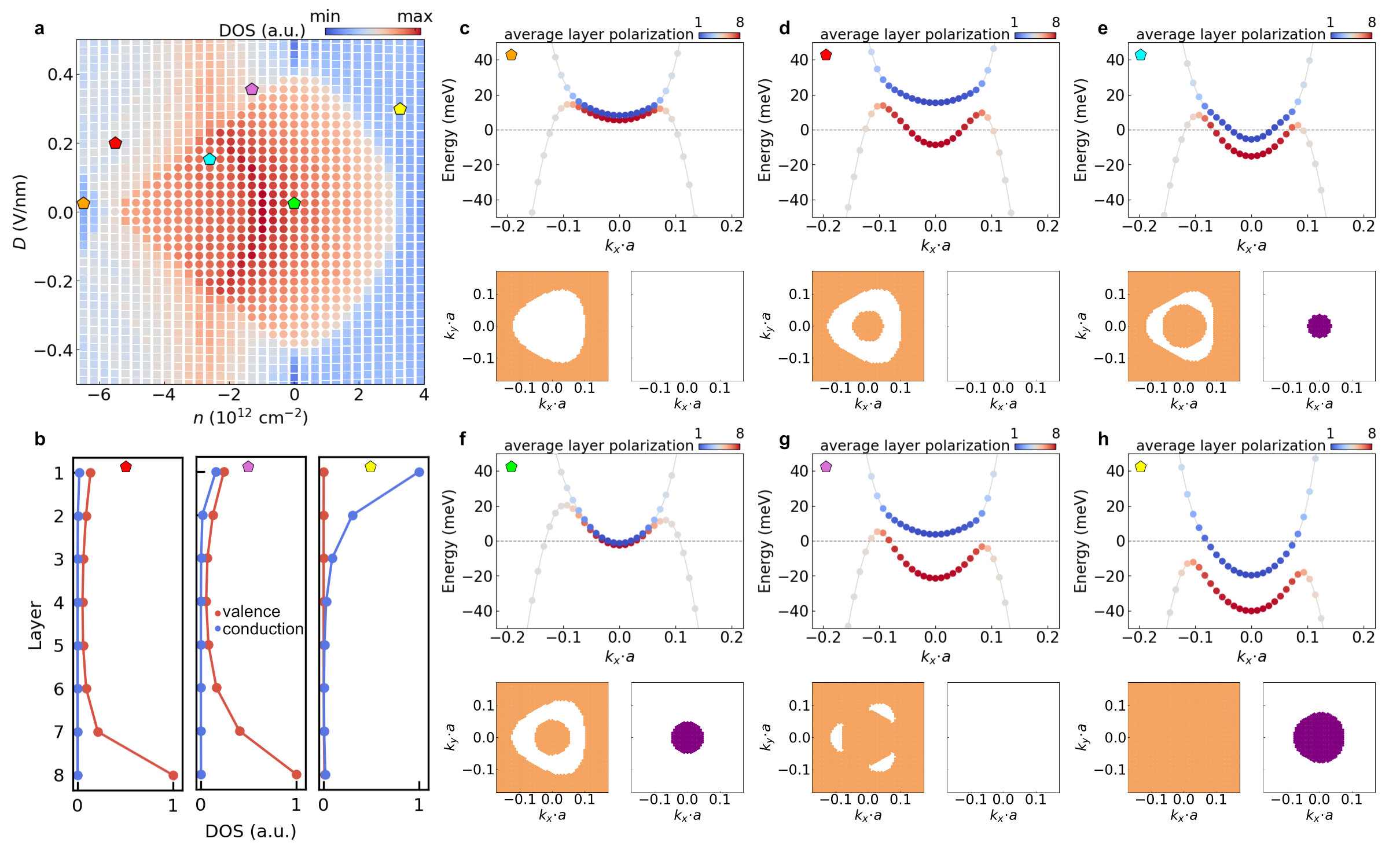} 
\caption{\textbf{Additional band structure calculations for misaligned R8G.}
\textbf{a}, Calculation of the DOS versus $n$ and $D$ in the Hartree-Fock approximation, reproduced from Fig.~\ref{fig:1}e.
\textbf{b}, Calculated DOS as a function of layer number for three different combinations of $n$ and $D$, corresponding to the positions of the color-coded markers in \textbf{a}. Note that for the data corresponding to the purple marker, there is an artificial conduction band tail due to numerical broadening. 
\textbf{c}, (Top) Hartree-Fock band structure corresponding to the $n$ and $D$ denoted by the orange marker in \textbf{(a)}. The color coding of points indicates their average layer polarization. (Bottom) Fermi surface contour plots at the Fermi energy ($E=0$), where white represents empty states and colors represent filled states. The valence band is on the left and the conduction band on the right.
\textbf{d-h}, The same as \textbf{(c)}, at the combinations of $n$ and $D$ indicated by the color-coded markers. 
}
\label{fig:8L_BS}
\end{figure*}

\newpage

\renewcommand{\figurename}{Supplementary Information Fig.}
\renewcommand{\thesubsection}{S\arabic{subsection}}
\setcounter{secnumdepth}{2}
\renewcommand{\thetable}{S\arabic{table}}
%\subsubsectionfont{\normalfont\large\itshape\underline}
%\renewcommand{\theequation}{S\arabic{equation}}
\setcounter{figure}{0} 
\setcounter{equation}{0}

\onecolumngrid
\newpage

\FloatBarrier           
\section*{Supplementary Information}
\FloatBarrier  

\section{Additional Experimental Analysis}

\begin{figure*}[h]
\includegraphics[width=0.8\textwidth]{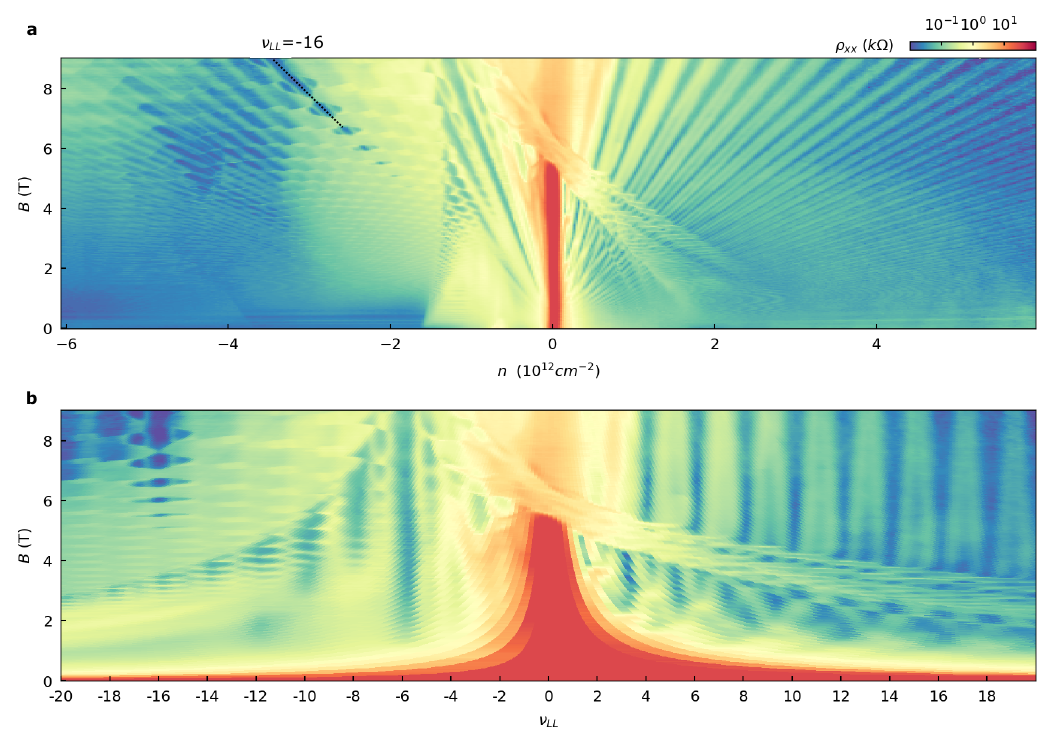} 
\caption{\textbf{Layer number determination of the misaligned R8G sample.}
\textbf{a}, Landau fan diagram taken at $D=0$.
\textbf{b}, The same data replotted versus Landau level filling factor ($\nu_{LL}$). There is a robust state at $\nu_{LL}=-16$ at large $B$ indicating the layer number is eight (see further discussion in the Methods).
}
\label{fig:R8G_layernum}
\end{figure*}

\begin{figure*}[h]
\includegraphics[width=0.8\textwidth]{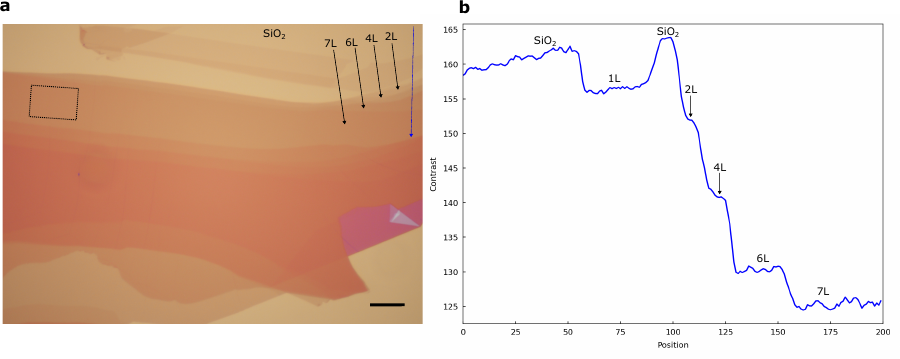} 
\caption{\textbf{Layer number determination of the moir\'e R7G sample.}
\textbf{a}, Optical micrograph of the exfoliated flake. The black dashed box shows the region of the flake from which the sample was made. The scale bar is 10~$\mu$m.
\textbf{b}, Line cut of the optical contrast taken along the position of the blue arrow in \textbf{a}. Position ``200'' corresponds to the arrow head. Contrast jumps of $\approx 5$ ($10$) indicate a monolayer (bilayer) step. Careful counting of steps indicates the flake is seven layers.
}
\label{fig:R7G_layernum}
\end{figure*}

\begin{figure*}[h]
\includegraphics[width=\textwidth]{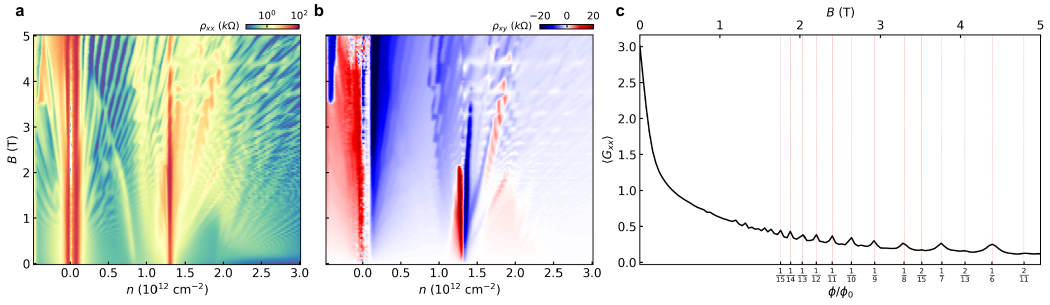} 
\caption{\textbf{Twist angle determination from Brown-Zak oscillations in the moir\'e R7G device.} 
\textbf{a}, Landau fan of $\rho_{xx}$ taken at $D=0.50$~V/nm.
\textbf{b}, Same Landau fan for $\rho_{xy}$.
\textbf{c}, Average conductance ($\langle G_{xx} \rangle$) plotted as a function of $B$, or equivalently the magnetic flux $\phi/\phi_0$, extracted from \textbf{(a)}. Brown-Zak oscillations are visible as sharp peaks in the curve. By fitting the oscillations, we extract the density corresponding to full filling of the moiré bands as $n_s = 2.55\times10^{12}$~cm$^{-2}$.
}
\label{fig:BrownZak}
\end{figure*}

\begin{figure*}[h]
\includegraphics[width=0.65\textwidth]{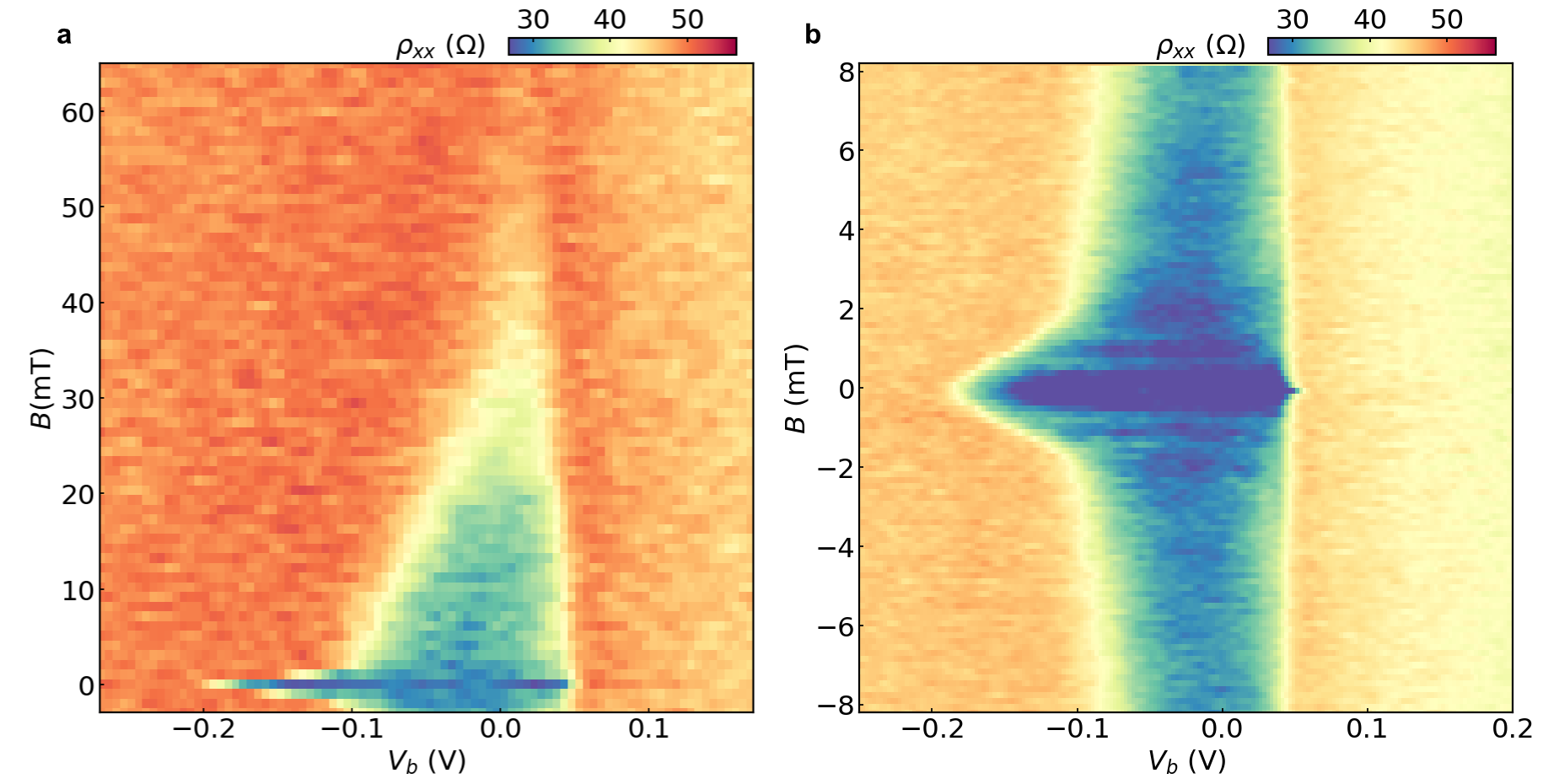} 
\caption{\textbf{Characterizing two separate behaviors of SC3$_8$ with $B$.}
\textbf{a}, Measurement of $\rho_{xx}$ versus $V_b$ and $B$ at fixed $V_t=-2.55$~V for the SC3$_8$ pocket. There is a sharp suppression of $\rho_{xx}$ very near $B=0$ corresponding to superconductivity, as well as a weaker suppression persisting up to $\approx 50$~mT. The latter likely corresponds to some effect other than superconductivity, although we do not know its origin. 
\textbf{b}, Zoom-in of $\rho_{xx}$ at fixed $V_t=-2.60$~V close to $B=0$ showing the evolution from a sharp suppression below a few millitesla to a weaker suppression at larger $B$.
}
\label{fig:SC3_Zoom}
\end{figure*}

\begin{figure*}[h]
\includegraphics[width=0.95\textwidth]{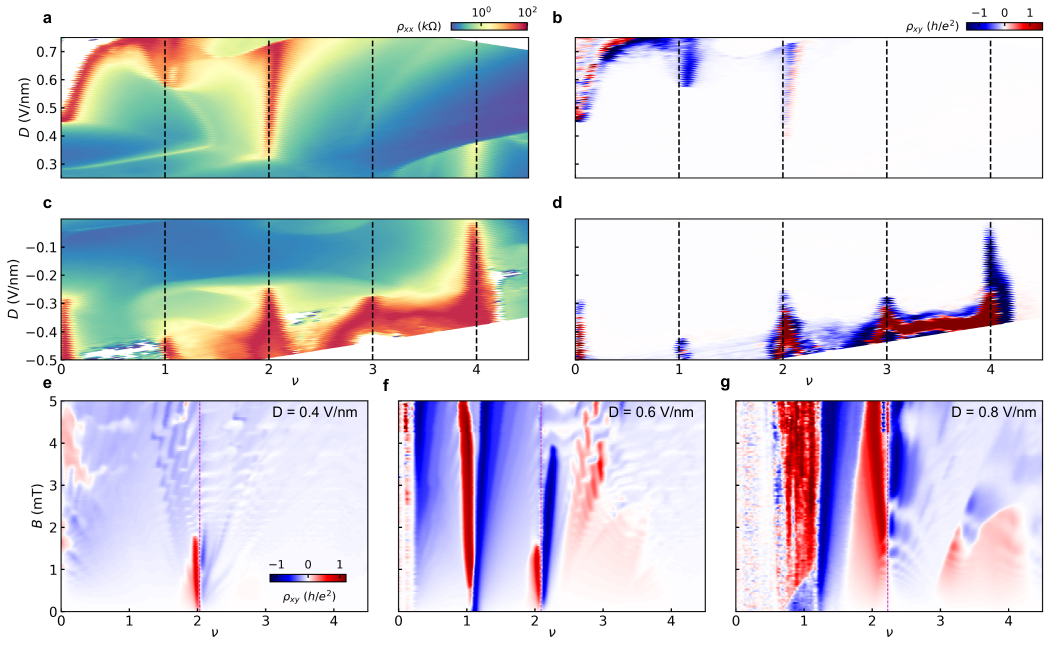} 
\caption{\textbf{Comparison of moir\'e band filling factors for moir\'e-adjacent and -remote polarization in the R7G device.}
\textbf{a}, Map of $\rho_{xx}$ versus $\nu$ and $D$ for $D>0$ at $B=0$. This corresponds to conduction electrons polarized to the moir\'e-remote interface.
\textbf{b}, Corresponding map of $\rho_{xy}$.
\textbf{c}, Same as \textbf{(a)} for $D<0$, in which conduction electrons are moir\'e-adjacent.
\textbf{d}, Same as \textbf{(b)} for $D<0$.
\textbf{e}, Landau fan of $\rho_{xy}$ taken at $D=0.4$~V/nm. The trivial correlated insulating state corresponding to half-filling is denoted by the vertical pink dashed line, and appears almost precisely at $\nu=2$.
\textbf{f}, Same at $D=0.6$~V/nm.
\textbf{g}, Same at $D=0.8$~V/nm.
As $D$ is raised, the position of the insulator drifts to values noticeably larger than $2$. In contrast, no such drifting behavior is seen for the moir\'e-adjacent side ($D<0$).
}
\label{fig:Dcomparison}
\end{figure*}

\begin{figure*}[h]
\includegraphics[width=\textwidth]{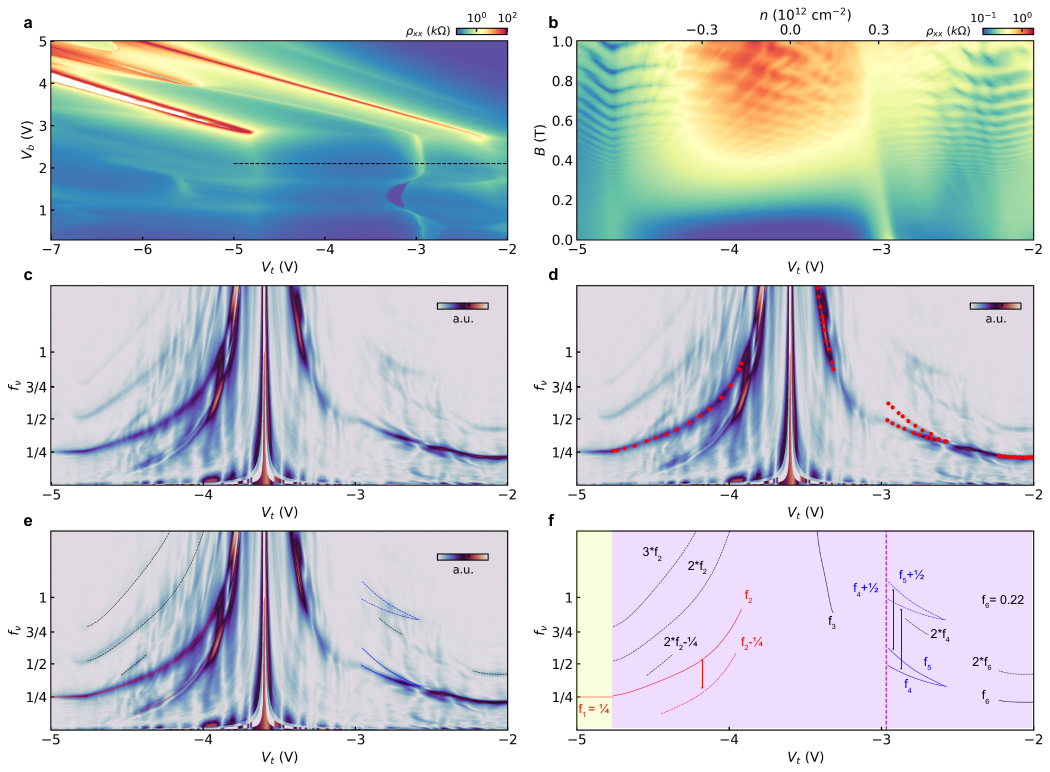} 
\caption{\textbf{Details of the Fermiology analysis procedure.}
\textbf{a}, Measurement of $\rho_{xx}$ versus $V_b$ and $V_t$ from the moir\'e R7G device.
\textbf{b}, Landau fan diagram taken along the black dashed line in \textbf{(a)}.
\textbf{c}, FFT of the Landau fan shown in \textbf{(b)}.
\textbf{d}, Red dots denote user-selected points on the interactive plot (see Methods for description).
\textbf{e}, Fits of the user-selected peaks (solid curves) and associated secondary frequencies (dashed curves, see Methods).
\textbf{f}, Schematic of the corresponding frequencies determined from the fits. Curves are color coded based on the corresponding implied Fermi surface degeneracy, with red corresponding to four, blue corresponding to two, and black corresponding to other/unknown.
Background color indicates band-isolated (yellow) and band-overlap regimes (purple).}
\label{fig:Fermiology7LVb2p1V}
\end{figure*}

\begin{figure*}[h]
\includegraphics[width=\textwidth]{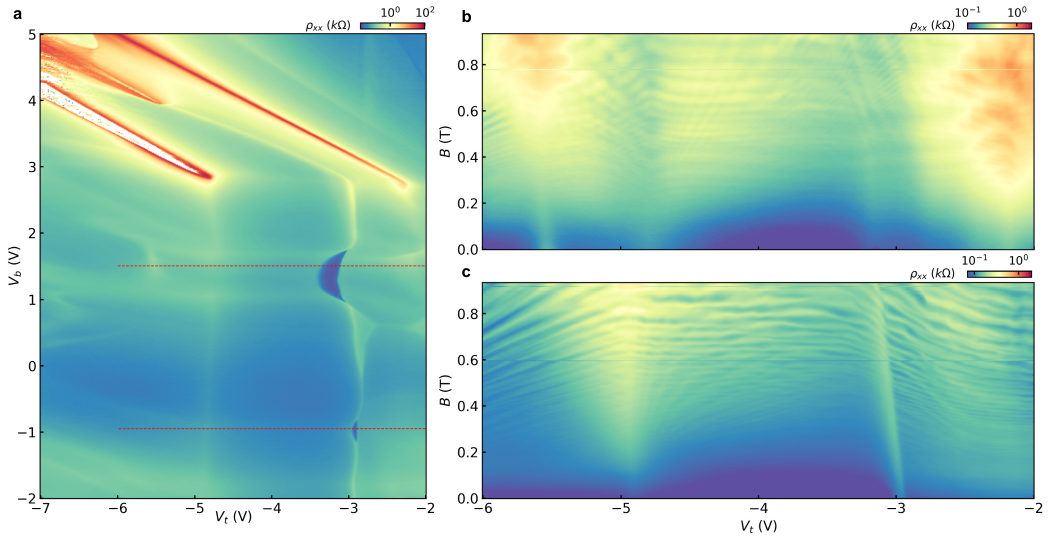} 
\caption{\textbf{Landau fans taken through SC1$_{7}$ and SC2$_{7}$.}
\textbf{a}, Map of $\rho_{xx}$ versus $V_b$ and $V_t$ at $B=0$.
\textbf{b}, Landau fan taken by sweeping $V_t$ at fixed $V_b=1.50$~V, corresponding to the upper red dashed line in \textbf{(a)}, cutting through SC1$_7$.
\textbf{c}, Landau fan taken by sweeping $V_t$ at fixed $V_b=-0.95$~V, corresponding to the lower red dashed line in \textbf{(a)}, cutting through SC2$_7$. The sharp resistive bump associated with SC2$_7$ drifts to the left with $B$, indicative of a spin-polarized state to the right gaining Zeeman energy over an unpolarized state to the left.
}
\label{fig:LFs_SC1_SC2}
\end{figure*}

\FloatBarrier
\clearpage 
\section{Additional Theoretical Analysis}

\subsection{Convergence Analysis}

\begin{figure}[h!]
    \centering
    \includegraphics[width=0.8\linewidth]{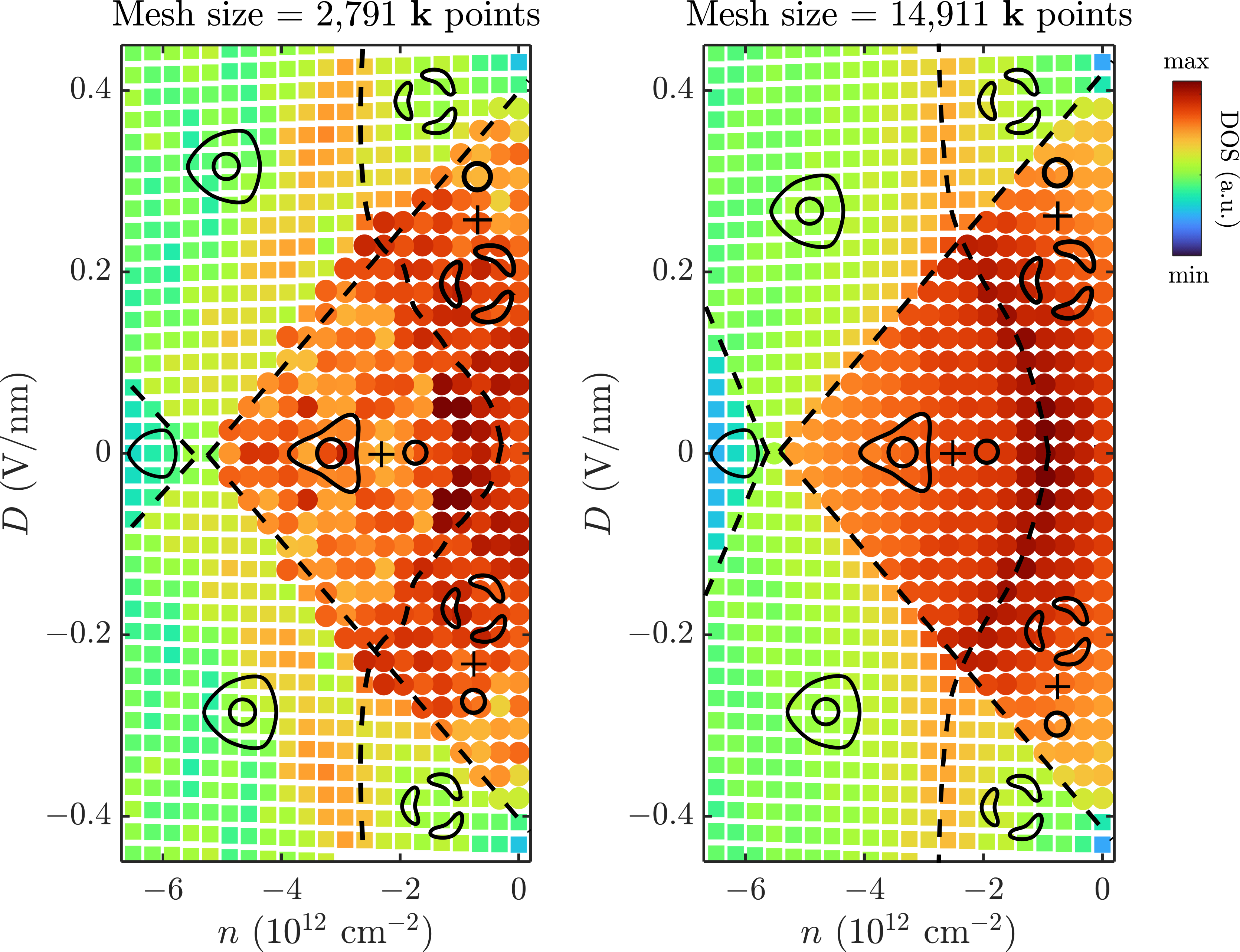}
    \caption{\textbf{Phase diagrams simulated with different mesh sizes.} The color map indicates DOS. Phase boundaries separating regions with different Fermi surface topologies are indicated by dashed curves. The Fermi surface structure within each region is schematically shown. For each point in phase space, a circle represents a system whose the Fermi surface comprises of states from both the valence and conduction bands, while a square in phase space represents a system whose Fermi surface consists of states only from the valence bands. }
    \label{fig:different_mesh_sizes}
\end{figure}

In this section, we present results for the phase diagram of octalayer graphene using different mesh sizes to confirm that the results presented in the main text have likely converged. In Fig. \ref{fig:different_mesh_sizes}, we show (left) the   phase diagram computed with a grid of $2,791$ points and (right)  the phase diagram computed with a grid of $14,911$ points (this is the same phase diagram shown in the main text). As is evident, the two phase diagrams qualitatively agree quite well. In particular, the crucial feature of overlapping Fermi pockets is robustly present in both phase diagrams, and it even persists in roughly the same region in parameter space. That the two mesh sizes differ significantly from each other (by more than five times) and yet still yield phase diagrams containing the main features of interest strongly suggests that our simulations have converged. Of course, there are some minor differences between the two phase diagrams. The phase diagram with the smaller grid does not demonstrate smooth evolution of DOS, which is to be expected since DOS is an integrated quantity over momentum space that becomes better approximated with increasing mesh size. Also, the boundary lines separating different Fermi surface topologies have different slopes between the two mesh sizes. 

\begin{figure}[h!]
    \centering
    \includegraphics[width=1\linewidth]{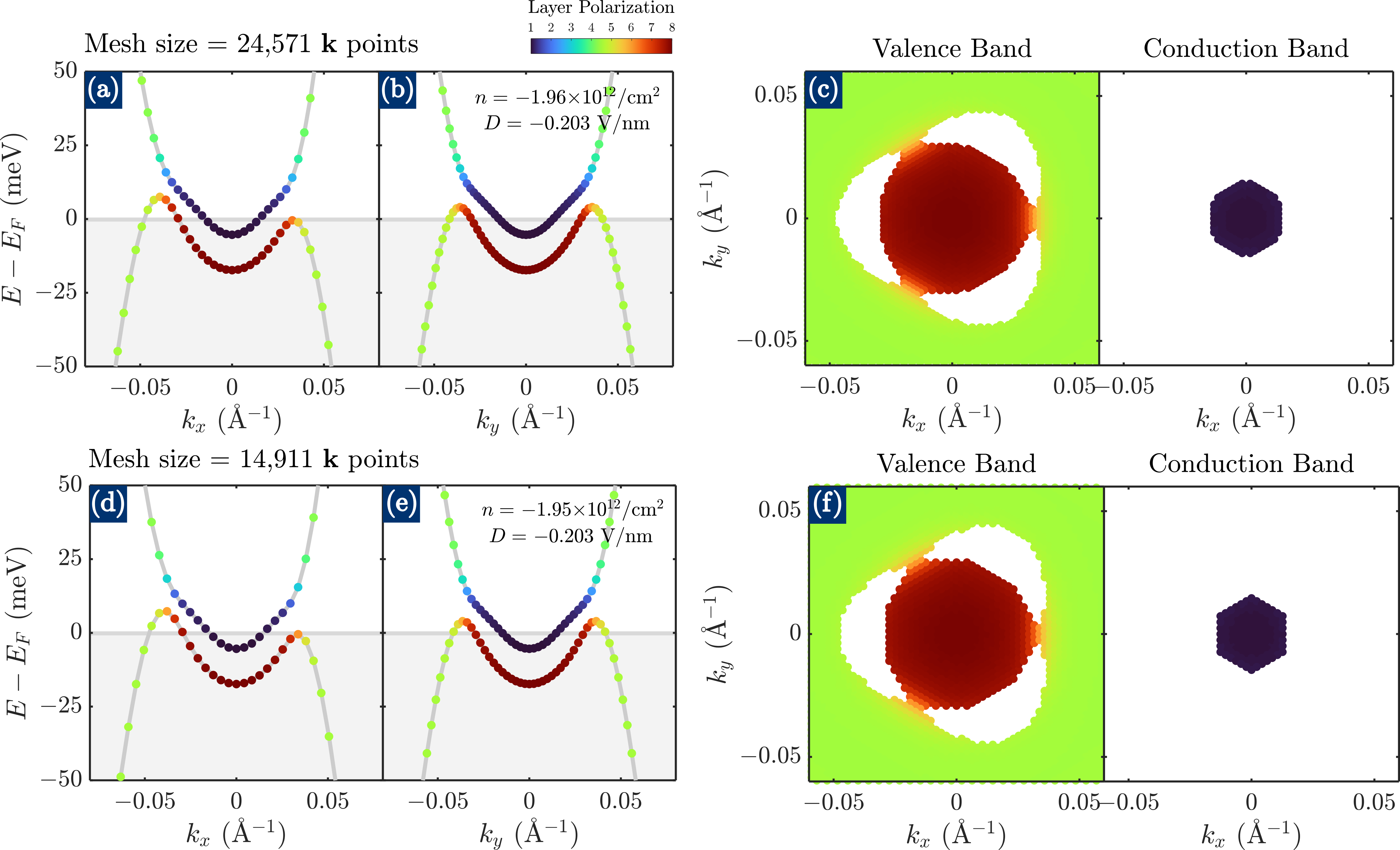}
    \caption{\textbf{Band structure and Fermi surface for a large mesh size.} (a)-(c) The grid contains $24,571$ points. Band structure along (a) $k_x$ and (b) $k_y$  and  (c) Fermi surface of the valence and conduction bands at $n = -1.96\times10^{12}$ cm$^{-2}$ and $D = -0.203$ V/nm. For comparison, we show the same plots at roughly the same density and displacement field for a grid size of  $14,911$ points in (d)-(f).}
    \label{fig:N=90}
\end{figure}

Furthermore, we also simulate the band structure and Fermi surface at a particular point in phase space (the point chosen is at one of the superconducting regions) for a much larger grid of $24,571$ points, shown in Fig. \ref{fig:N=90}. As can be seen here, all qualitative features match those in the $14,911$-point grid. Although we cannot simulate the entire phase diagram with the $24,571$-point grid because of limited computational resources, the agreement at this point between the different grid sizes adds more confidence to our numerical simulations. 

\subsection{Comparison of Different Approximations}

In this section, we compare the phase diagrams numerically computed (1) without electron-electron interactions, (2) with electron-electron interactions that neglect the layer index, and (3) with electron-electron interactions that include layer charge screening. A few remarks about each approximation are in order:
\begin{enumerate}
    \item \textbf{Non-interacting limit:} The first approximation operates with just the non-interacting band structure (which presumably  does account for some effects from Coulomb interactions in \textit{ab initio} calculations at charge neutrality and zero displacement field) and assume that it does not deform as a function of filling, i.e. the bands are rigid against density variation. In this framework, the graphene stack experiences the completely \textit{unscreened} external displacement field via an \textit{ad hoc} uniformly progressive on-site potential energy $\Delta.$ In this approximation, there is generally no consistent way to relate $\Delta$ to the displacement field $D$ because the charges are not determined self-consistently (note that, in our notation, the displacement field is dimensionally equivalent to the electric field, which is contrary to the convention used in many introductory electrodynamics textbooks wherein $[D] = [\epsilon_0][E],$ where $E$ is the electric field.). One might naively use $D \stackrel{?}{=} \epsilon_r\Delta/ed,$ but this choice generally violates Gauss's law. To see this, we observe that since the potential drop between any two layers is the same $(=\Delta/e),$ the displacement field between any two layers must be the same. Gauss's law therefore enforces for $\ell \in \lbrace 2,3,...,N_\ell-1\rbrace$
    \begin{equation}
        D_{\ell,\ell+1}- D_{\ell-1,\ell} = \frac{\sigma_\ell}{\epsilon_0} =0.
    \end{equation}
    However, our calculation does not demand that the inner layers host no excess charge. In fact, there are generally excess charges when we vary the density.  Without knowing the interlayer displacement fields, we cannot determine the experimental displacement field $D.$ Therefore, in this framework, we treat $\Delta$ as fundamental and forgo any attempt to connect it back to $D.$
    \item \textbf{Layerless interacting limit:} The second approximation accounts for electron-electron interactions in the mean-field limit as described in the Methods section of the main text except that here, we  use the layer-independent, dual-gated Coulomb potential
    \begin{equation}
        \mathcal{V}(\mathbf{q}) = \frac{e^2}{2 \epsilon_r \epsilon_0|\mathbf{q}|} \tanh \left( \frac{d_g |\mathbf{q}|}{2} \right).
    \end{equation}
    We note that this layer-independent potential is equivalent to 
    \begin{equation}
\begin{split}
 \label{eq: Coulomb potential}
    \mathcal{V}(\mathbf{q}) = \lim_{z_{\ell},z_{\ell'}\rightarrow d_g/2 }\frac{e^2\mathrm{csch} \left( |\mathbf{q}| d_g\right)}{2 \epsilon_0 \epsilon_r|\mathbf{q}|}   &\left[ \cosh \left(|\mathbf{q}|\{d_g-|z_\ell-z_{\ell'}|\}\right)- \cosh \left(|\mathbf{q}|\{d_g-|z_\ell+z_{\ell'}|\}\right)  \right].   
\end{split}
\end{equation}
In this approximation, all the charges behave as though they reside on the same plane. They can still interact among each other, the effects of which can deform the energy bands. However, there is no out-of-plane electron-electron interactions between the layers that screen the external displacement field. In fact, since the layers are regarded as being on the same plane, there is no meaning to an interlayer displacement field. Thus, this approximation, like the non-interacting limit, does not allow us to consistently determine $D.$ As such, $\Delta$ remains the fundamental knob with which to effectuate  the displacement field. 

\item \textbf{Layerful interacting limit:} This last approximation accounts for electron-electron interactions in the mean-field limit with both gate screening and layer-charge screening as described in the Methods section of the main text. In this approximation, the layer-dependent Hartree term allows us to faithfully convert between the experimental displacement field $D$ and the phenomenological on-site potential energy $\Delta$ using charges on the layers obtained self-consistently. This approximation represents our most accurate attempt to theoretically reproduce the experimental results in this work. 

\begin{figure}[h!]
    \centering
    \includegraphics[width=1\linewidth]{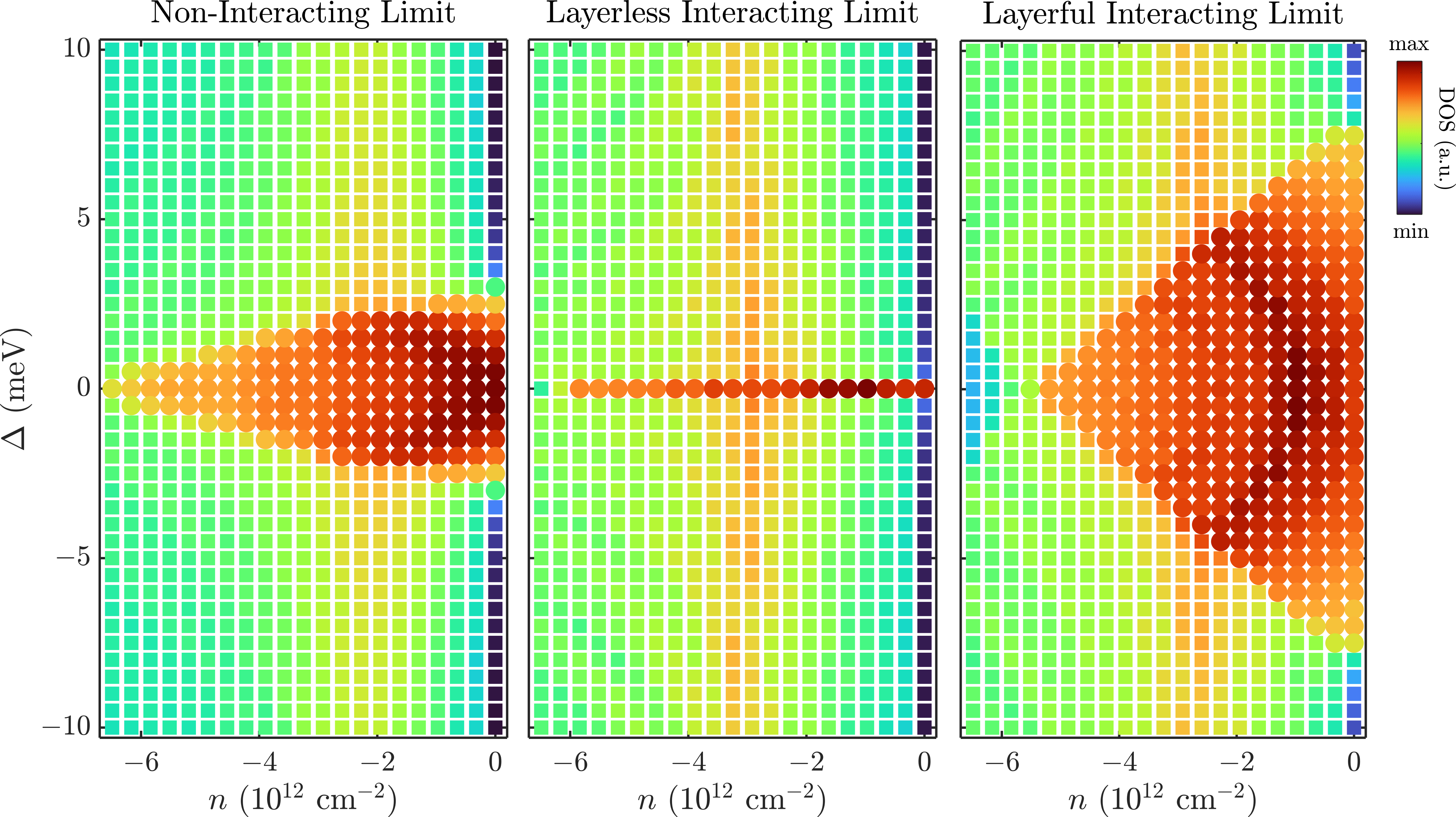}
    \caption{\textbf{Comparison of phase diagrams under different approximations.} A circle in phase space represents a system whose the Fermi surface comprises of states from both the valence and conduction bands, while a square in phase space represents a system whose Fermi surface consists of states only from the valence bands. The color scale indicates the DOS at the Fermi surface. As is clear, compared to the non-interacting limit, the layerless interacting limit suppresses the regime of coexisting Fermi pockets, while the layerful interacting limit enhances it.  }
    \label{fig:comparison_phase_diagram}
\end{figure}

\begin{figure}[h!]
    \centering
    \includegraphics[width=1\linewidth]{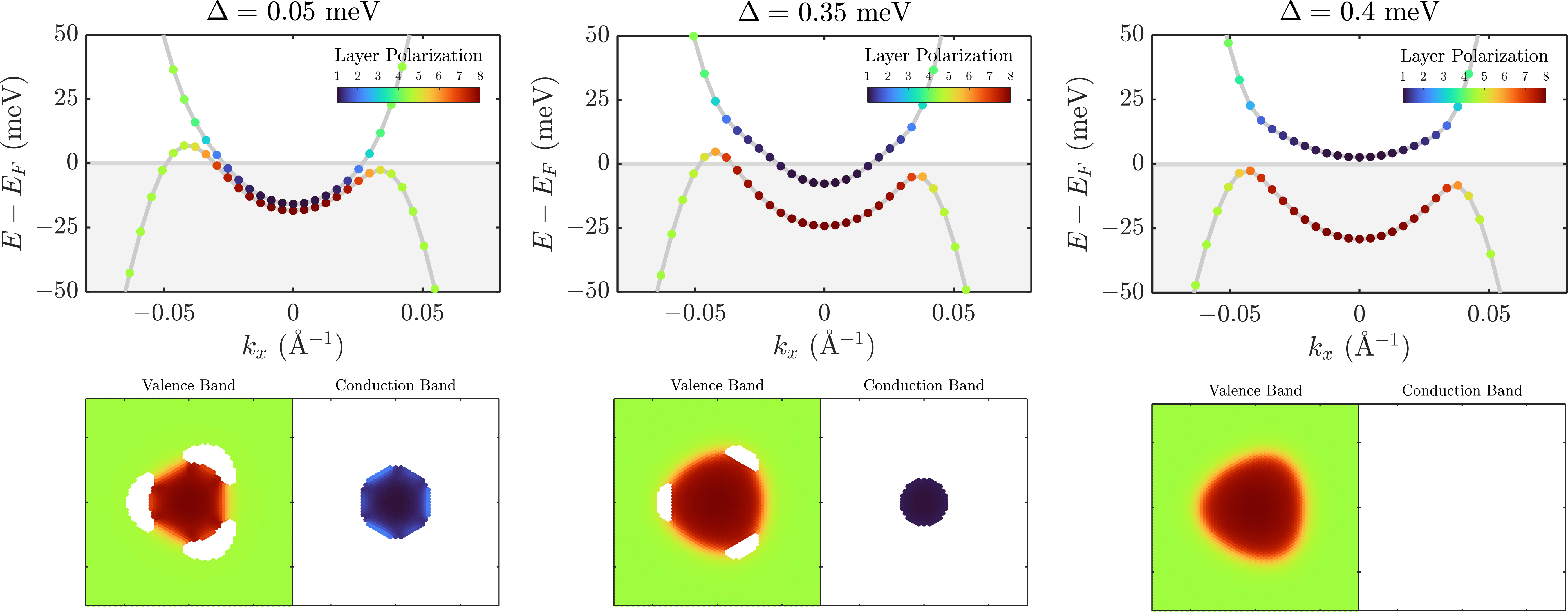}
    \caption{\textbf{Band structures in the layerless interacting limit for different values of on-site potentials $\Delta$.} The band structures are shown for a single valley and spin. Below each band structure, we show the corresponding Fermi sea resolved into the valence and conduction bands.     }
    \label{fig:band_structure_layerless}
\end{figure}

\end{enumerate}

It should be noted in all three approximations described above, internal screening, a purely quantum effect  captured, for instance, in the random phase approximation or some other beyond-mean-field method, is not systematically investigated in this work. As in the main text, all simulation results reported here use $\epsilon_r = 16,$  $d_g = 400$ \AA, and $N_\mathbf{k}=70$ unless otherwise noted.

The phase diagrams computed in the three approximations are shown in Fig. \ref{fig:comparison_phase_diagram}. Compared to the non-interacting limit, the phase diagram computed with a layer-independent Coulomb potential energy barely features any regime of coexisting Fermi surface from the valence and conduction bands. Two simultaneous effects contribute to this outcome: (1) without the Hartree term, there is no reduction of the input displacement field and (2) the Fock term generally favors gap enhancement, which disfavors coexisting Fermi pockets from different bands.  Therefore, while the coexistence regime does exist, as shown in Fig. \ref{fig:band_structure_layerless}, it is significantly quenched for the choice of realistic parameters we are using in the layerless interacting limit. On the other hand, the phase diagram simulated with a layer-dependent Coulomb potential energy enlarges the phase-space volume of the regime of coexisting Fermi surface. This is because the layer-dependent Coulomb potential screens the external displacement field with an induced interlayer displacement field. Since the \textit{net} displacement field controls the gap between the valence and conduction bands, interlayer screening therefore favors the coexistence regime. The semi-quantitative agreement between the experimental phase diagram and the theoretical phase diagram in the layerful interacting limit illustrates the crucial role of interlayer screening in faithfully capturing the underlying physics in this experiment.

\subsection{Phase Diagram for Electron Doping}

\begin{figure}
    \centering
    \includegraphics[width=1\linewidth]{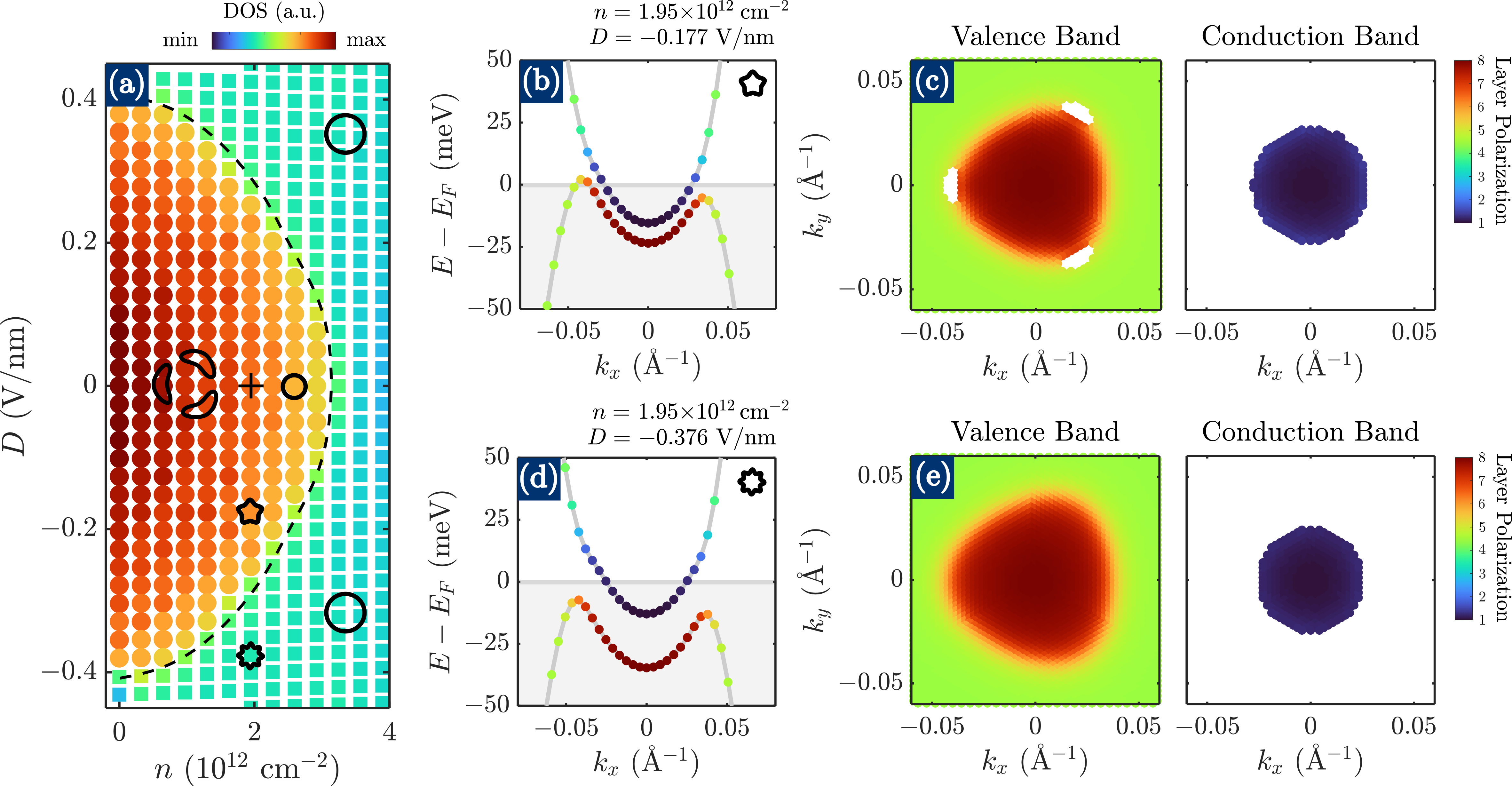}
    \caption{\textbf{Phase diagram for  electron dopings.} (a) Phase diagram as a function of positive filling $n\geq0$ and displacement field with an approximate phase boundary drawn with a dashed curve delineating regions of different Fermi surface structures. The Fermi surface topology for each region is schematically shown. (b)-(c) Band structure and Fermi sea, color coded by layer polarization, at $n = 1.95\times10^{12}$ cm$^{-2}$ and $D = -0.177$ V/nm, showing a Fermi surface consisting of states from both the conduction and valence bands. (d)-(e) Band structure and Fermi sea, color coded by layer polarization, at $n = 1.95\times10^{12}$ cm$^{-2}$ and $D = -0.376$ V/nm, showing a Fermi surface consisting of states only from the conduction band. }
    \label{fig:electron_side}
\end{figure}

In this section, we report the phase diagram for the electron-doped side of misaligned octralayer graphene, as shown in Fig. \ref{fig:electron_side}. Compared to the remarkable agreement between theory and experiment for the hole-doped side, the agreement between theory and experiment for the electron-doped side is more subtle. There are many resistive features in the experiment for positive $n$ that are not immediately apparent in numerical results. We strongly suspect these differences originate from some symmetry breaking mechanisms that are not presently accounted for in the simulation. Nevertheless, even without accounting for spontaneous symmetry breaking, we are able to qualitatively reproduce the phase boundary that separates regions of different Fermi surface topologies for the electron-doped side, shown as a dashed curve in Fig. \ref{fig:electron_side}(a). The region at small dopings and displacement fields consists of Fermi surfaces that have pockets from both the valence and conduction bands, as shown in Fig. \ref{fig:electron_side}(b,c) . This region is adiabatically connected to the region in the hole-doped side that hosts superconductivity. Outside this region, the Fermi surface only consists of states from the conduction band and has simply-connected topology, as shown in Fig. \ref{fig:electron_side}(d,e).

\subsection{Hartree-Fock Theory for Moir\'{e} Structures}

In this section, we present our Hartree-Fock calculations for the hBN-aligned heptalayer system. The formation of a moir\'{e} structure due to alignment with hBN requires us to construct a continuum model that describes the non-interacting bands renormalized by (virtual) hoppings between graphene and hBN. We adopt the formalism from Ref. \cite{Moon2014Electronic}. We consider an $N_\ell$ rhombohedral graphene stack initially aligned with a hBN sheet underneath in perfect crystallographic alignment (i.e. angle mismatch is zero). We denote the graphene lattice vectors as $\mathbf{a}_1 = a\hat{e}_x$ and $\mathbf{a}_2 = a/2 \hat{e}_x+\sqrt{3}a/2 \hat{e}_y,$ where $a = 2.46$ \AA $\mathrm{ }$ is the graphene lattice constant \cite{Castro2009The}. When hBN is aligned with the graphene stack, the hBN lattice vectors are the same as the graphene lattice vectors with the appropriate lattice constants replaced: $ 1+\delta_a = a_\mathrm{hBN}/a = 1.017$ \cite{Kern1999Ab,Moon2014Electronic}. Now, we rotate the hBN layer by an angle $\theta$ implemented by a counterclockwise rotation matrix $\mathcal{R}(\theta)$. This generates a moir\'{e} pattern with lattice vectors given by $\mathbf{L}_i =  \left(\left[\left( 1 + \delta_a \right)\mathcal{R}(\theta) - \mathbb{1}\right]^{-1} + \mathbb{1} \right)\mathbf{a}_{i},$ where the moir\'{e} lattice constant is $L_M  = |\mathbf{L}_i| = a \left(1+ \delta_a \right) \left[\delta_a^2+2(1+\delta_a)(1-\cos \theta ) \right]^{-\frac{1}{2}} \approx a(1+\delta_a) \left[\delta_a^2+(1+\delta_a)\theta^2 \right]^{-\frac{1}{2}}.$ The moir\'{e} reciprocal lattice vectors are given by $ \mathbf{G}_i = \left( \mathbb{1}- (1+\delta_a)^{-1} \mathcal{R}(\theta) \right)\mathbf{g}_{i},$ where  $\mathbf{g}_{1} = \frac{4\pi}{\sqrt{3} a} \hat{e}_y$ and $\mathbf{g}_{2} = -\frac{2\pi}{ a}\hat{e}_x-\frac{2\pi}{\sqrt{3} a}\hat{e}_y.$ In general, the Hamiltonian can be written as
\begin{equation}
    \mathbb{K} = \begin{pmatrix}
        \mathbb{K}_{N_\ell} & \mathbb{U} \\
        \mathbb{U}^\dagger & \mathbb{K}_\mathrm{hBN}
    \end{pmatrix},
\end{equation}
where $\mathbb{U}$ is the interlayer hopping matrix that describes hopping between the hBN sub-Hamiltonian and $\ell=1$ of the graphene sub-Hamiltonian, and the hBN sub-Hamiltonian, neglecting any dispersion, is taken to be
\begin{equation}
    \mathbb{K}_\mathrm{hBN} = \begin{pmatrix}
        \varepsilon_\mathrm{A} & 0\\
        0 & \varepsilon_\mathrm{B}
    \end{pmatrix},
\end{equation}
where $\lbrace \varepsilon_\mathrm{A},\varepsilon_\mathrm{B} \rbrace \in \lbrace \varepsilon_\mathrm{Nitrogen},\varepsilon_\mathrm{Boron} \rbrace = \lbrace -1493,3332\rbrace$ meV \cite{2010Energy,Kindermann2010Zero,Jung2014Ab}. The hopping matrix in valley $\xi$ is
\begin{equation}
    \mathbb{U} = \begin{pmatrix}
        u_{\mathrm{A} } & u_{\mathrm{B} } \\
        u_{\mathrm{A} } & u_{\mathrm{B} }
    \end{pmatrix} + \begin{pmatrix}
        u_{\mathrm{A} } & u_{\mathrm{B} } \omega^{\xi} \\
        u_{\mathrm{A} }\omega^{-\xi} & u_{\mathrm{B} }
    \end{pmatrix} e^{-i\xi \mathbf{G}_2 \cdot \mathbf{r}} + \begin{pmatrix}
        u_{\mathrm{A} } & u_{\mathrm{B} } \omega^{-\xi} \\
        u_{\mathrm{A} }\omega^{\xi} & u_{\mathrm{B} }
    \end{pmatrix} e^{-i\xi \left( \mathbf{G}_1 + \mathbf{G}_2\right) \cdot \mathbf{r}},
\end{equation}
where $\omega = e^{2\pi i /3}$ and $\lbrace u_\mathrm{A},u_\mathrm{B} \rbrace \in \lbrace u_\mathrm{Nitrogen},u_\mathrm{Boron} \rbrace = \lbrace 97,144\rbrace$ meV \cite{Jung2014Ab}. Because we are interested in physics induced on graphene by hBN, we can project to the graphene subspace using second-order perturbation theory
\begin{equation}
\label{eq: moire Hamiltonian}
    \mathbb{K} = \mathbb{K}_{N_\ell} - \mathbb{U} \mathbb{K}_\mathrm{hBN}^{-1} \mathbb{U}^\dagger,
\end{equation}
where
\begin{equation}
    \mathbb{V}_\mathrm{hBN} = - \mathbb{U} \mathbb{K}_\mathrm{hBN}^{-1} \mathbb{U}^\dagger = \mathcal{V}_1\begin{pmatrix}
        1 & 0 \\
        0 & 1
    \end{pmatrix} + \mathcal{V}_2 e^{i\xi \phi}  \left[\begin{pmatrix}
        1 & \omega^\xi \\
        \omega^{\xi} & \omega^{-\xi}
    \end{pmatrix} e^{-i\xi \mathbf{G}_1 \cdot \mathbf{r}} +  \begin{pmatrix}
        1 & 1 \\
        \omega^{-\xi} & \omega^{-\xi}
    \end{pmatrix} e^{-i\xi \mathbf{G}_2 \cdot \mathbf{r}}+  \begin{pmatrix}
        1 & \omega^{-\xi} \\
        1 & \omega^{-\xi}
    \end{pmatrix} e^{i\xi \left( \mathbf{G}_1+\mathbf{G}_2 \right)\cdot \mathbf{r}} \right] + \text{h.c.}
\end{equation}
\begin{equation}
    \mathcal{V}_1 = -3 \left( \frac{ u_{\mathrm{A}}^2}{\varepsilon_{\mathrm{A}}}+\frac{ u_{\mathrm{B}}^2}{\varepsilon_{\mathrm{B}}}\right) \quad \text{and} \quad \mathcal{V}_2 e^{i \xi \phi} = -\left(\frac{u_{\mathrm{A}}^2}{\varepsilon_{\mathrm{A}}}+\frac{u_{\mathrm{B}}^2 \omega ^{\xi }}{\varepsilon_{\mathrm{B}}}\right).
\end{equation}
There are two inequivalent classes of structures depending on the pretwist alignment: whether the boron atoms of hBN are aligned with the $A$ or $B$ sublattice of the adjacent graphene layer. We call a structure Type I if  $(\varepsilon_\mathrm{A},\varepsilon_\mathrm{B},u_\mathrm{A},u_\mathrm{B}) = (\varepsilon_\mathrm{Nitrogen},\varepsilon_\mathrm{Boron},u_\mathrm{Nitrogen},u_\mathrm{Boron})$ and Type II if  $(\varepsilon_\mathrm{A},\varepsilon_\mathrm{B},u_\mathrm{A},u_\mathrm{B}) = (\varepsilon_\mathrm{Boron},\varepsilon_\mathrm{Nitrogen},u_\mathrm{Boron},u_\mathrm{Nitrogen}).$ It is currently not possible to ascertain experimentally whether our moir\'{e} sample belongs to Type I or Type II. Therefore, we check both types to confirm that our conclusions are independent of this microscopic arrangement.

To account for electron-electron interactions in the self-consistent mean-field approximation, we project the full Hamiltonian, which includes the non-interacting Hamiltonian in Eq. \eqref{eq: moire Hamiltonian} and the two-body Coulomb interaction, onto the basis of energy bands of the non-interacting Hamiltonian. The two-body interaction Hamiltonian uses the same layer-dependent Coulomb potential energy as detailed in the Methods section with $\epsilon_r = 16$. We also implement a background subtraction of a uniform charge density from the density matrix. For numerical efficiency, we only keep the 6 bands closest to charge neutrality for each spin-valley flavor. The mesh size of the moir\'{e} Brillouin zone is $18\times 18$. The method is explained in more details in Ref. \cite{phongC2}. As with the misaligned calculations, we only compute the symmetric states renormalized by electron-electron interactions. 

\begin{figure}
    \centering
    \includegraphics[width=1\linewidth]{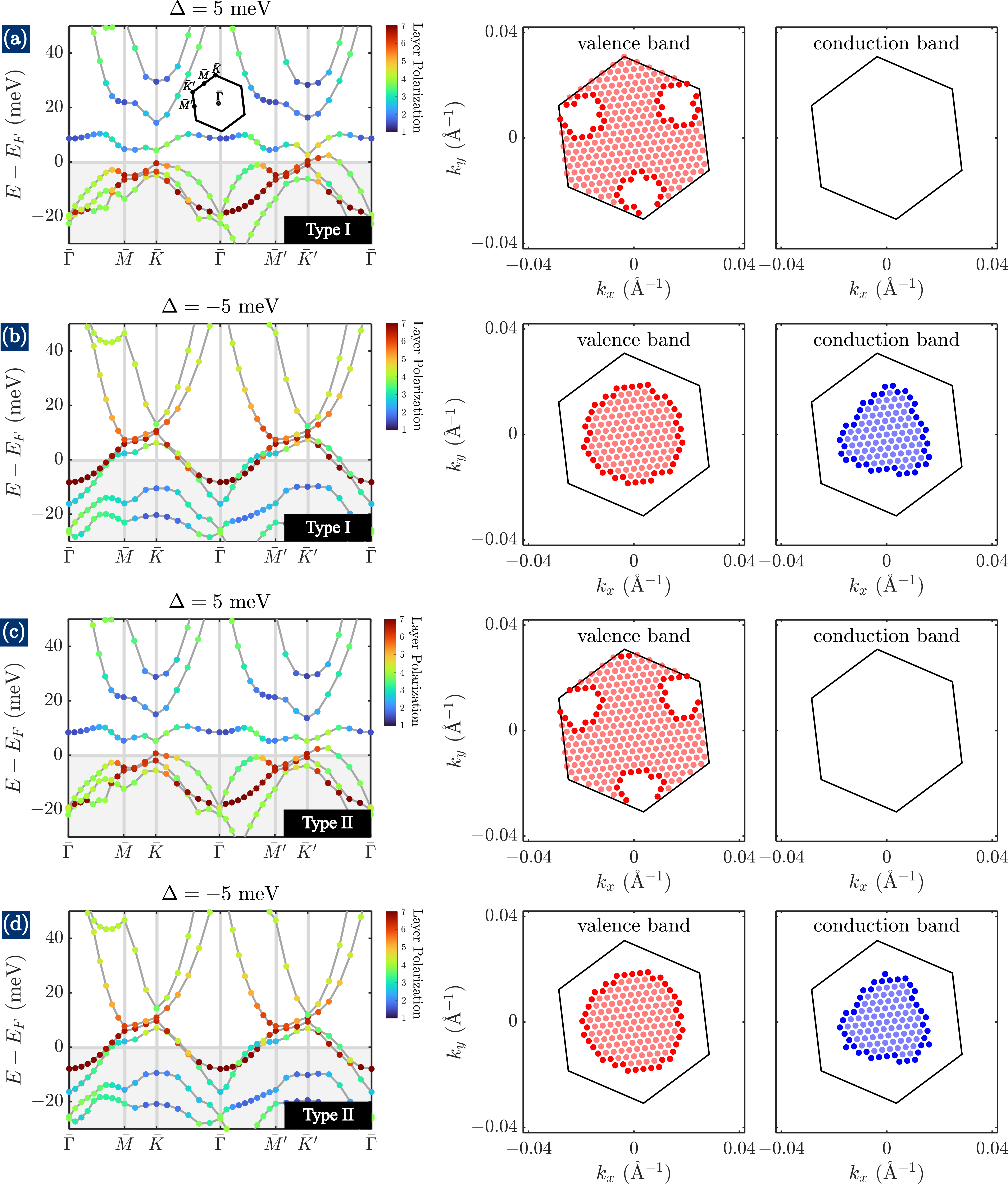}
    \caption{\textbf{Hartree-Fock band structures for moir\'{e} superlattices at $\theta =0.42^\circ$ and $\nu = -2/3$.} Band structures traced along high-symmetry lines in the moir\'{e} Brillouin zone as shown in the inset in (a) for Type I structures at positive $\Delta > 0$ (a) and negative $\Delta < 0$ (b) and for Type II structures at positive $\Delta > 0$ (c) and negative $\Delta < 0$ (d). Because of time-reversal symmetry, we only show the bands from the $\xi =+1$ valley. The energy eigenstates are color coded by their average layer polarization. For each band structure, we show the associated Fermi sea from the first valence band and conduction in the center and rightmost columns. The moir\'{e} Brillouin zone is constructed from $\mathbf{G}_1$ and $\mathbf{G}_2$ and is generically tilted for general $\theta.$  }
    \label{fig:moire}
\end{figure}

We show the single-particle band structures dressed by the Coulomb self-energy computed in the mean-field limit described above in Fig. \ref{fig:moire} at $\theta = 0.42^\circ,$ hole doping $\nu=-2/3,$ and $|\Delta|=5$ meV. We do not attempt to relate $D$ to $\Delta$ for moir\'{e} structures, but the chosen values of $\Delta$ and $\nu$ are roughly in the region of parameter space where superconductivity is experimentally observed. For $\Delta < 0,$ valence electrons at small momenta are primarily localized on $\ell=1$ and conduction electrons at small momenta are localized on $\ell = 7.$ In this case, the valence electrons feel the effect of the moir\'{e} structure from hBN alignment, but the conduction electrons are mostly unaffected by the moir\'{e} structure since they are spatially distant. For both Types I and II, we find that this \textit{sign} of $\Delta$ is favorable to the existence of coexisting Fermi pockets from valence and conduction bands, as shown in Fig. \ref{fig:moire}(b,d). For $\Delta > 0,$ the situation is reversed. In this case, valence electrons at small momenta are mostly localized on $\ell = 7,$ far away from hBN, while conduction electrons at small momenta are significantly by the moir\'{e} structure since they are mostly localized on $\ell = 1$. For $\Delta>0,$ we find that the significant renormalization of the conduction bands due to the moir\'{e} structure disfavors Fermi pocket coexistence, as shown in Fig. \ref{fig:moire}(a,c). This is consistent with the experimental observations where superconductivity is only observed at hole dopings for the same sign of $D$ where the integer quantum Hall effect is observed at electron dopings. This sign of $D$ corresponds to $\Delta < 0$ in our theoretical calculations. Therefore, the theoretical calculation supports the general conclusion that coexistence of Fermi pockets is important for Cooper pairing in these systems. We should note that for the parameters we have chosen, we are not able to generate the $\mathcal{C}=1$ bands at $\nu = 1$ that would presumbly give rise to the integer quantum Hall effect seen nearby at $\nu \approx  1.05$. We suspect that either a different set of parameters is needed (we note however that for the same set of parameters we robustly reproduce presence of $\mathcal{C}=1$ bands in moiré-aligned R5G case), or additional qualitative theoretical considerations are required (such as a more complex symmetry-breaking order parameter), or both. However, the basic conclusion of coexisting Fermi pockets appears quite robust against changes in parameters. Consequently, we are confident that this particular feature of the theory--coexisting Fermi pockets--will survive in a more refined theory.

\end{document}